\documentclass[twocolumn, prl, aps, superscriptaddress, longbibliography, showpacs, amsmath, amssymb, floatfix]{revtex4-2}     
\usepackage{changes}
\usepackage{graphicx}
\usepackage{dcolumn}
\usepackage{bm}
\usepackage{graphicx}                                
\usepackage{amssymb}
\usepackage{amsmath}
\usepackage{epsfig}
\usepackage{xcolor}
\usepackage{tabu}
\usepackage{mathtools}
\usepackage[colorlinks,linkcolor=red,anchorcolor=blue,citecolor=blue,urlcolor=blue]{hyperref}
\usepackage{physics}
\usepackage{float}
\usepackage{diagbox}
\usepackage{inputenc}
\usepackage{soul}
\usepackage{xspace}
\usepackage{mathdots}

\newcommand{\addrLab}{\affiliation{Laboratory of Quantum Information, University of Science and Technology of China, Hefei 230026, China}}
\newcommand{\addrAnhui}{\affiliation{Anhui Province Key Laboratory of Quantum Network,
University of Science and Technology of China, Hefei 230026, China}}

\newcommand{\addrSyn}{\affiliation{Synergetic Innovation Center of Quantum Information and Quantum Physics, University of Science and Technology of China, Hefei 230026, China}}

\newcommand{\addrSuzhou}{\affiliation{Suzhou Institute for Advanced Research, University of Science and Technology of China, Suzhou, Jiangsu 215123, China}}
\newcommand{\addrOrigin}{\affiliation{Origin Quantum, Hefei, Anhui 230088, China}}

\newcommand{\kbra}[2]{\ket{#1}\bra{#2}}
\renewcommand{\norm}[1]{\left| #1 \right|}

\newcommand{\beginsupplement}{%
    \setcounter{section}{0}
    \setcounter{subsection}{0}
    \setcounter{table}{0}
    \renewcommand{\thetable}{S\arabic{table}}%
    \setcounter{figure}{0}
    \renewcommand{\thefigure}{S\arabic{figure}}%
    \setcounter{equation}{0}
    \renewcommand{\theequation}{S\arabic{equation}}%
    \setcounter{section}{0}
    \renewcommand{\thesection}{\arabic{section}}%
}

\begin{document}

\title{Remote entanglement generation via enhanced quantum state transfer}
\author{Tian-Le Wang}
\addrLab\addrSyn
\author{Peng Wang}
\addrLab\addrSyn\addrSuzhou
\author{Ze-An Zhao}
\addrLab\addrSyn
\author{Sheng Zhang}
\addrLab\addrSyn\addrSuzhou
\author{Ren-Ze Zhao}
\author{Xiao-Yan Yang}
\author{Hai-Feng Zhang}
\author{Zhi-Fei Li}
\author{Yuan Wu}
\addrLab\addrSyn

\author{Liang-Liang Guo}
\author{Yong Chen}
\author{Hao-Ran Tao}
\author{Lei Du}
\author{Chi Zhang}
\author{Zhi-Long Jia}
\author{Wei-Cheng Kong}
\addrOrigin

\author{Peng Duan}
\email{pengduan@ustc.edu.cn}
\addrLab\addrSyn

\author{Ming Gong}
\email{gongm@ustc.edu.cn}
\addrLab\addrSyn
\addrAnhui

\author{Guo-Ping Guo}
\email{gpguo@ustc.edu.cn}
\addrLab\addrSyn\addrOrigin

\date{\today}

\begin{abstract}
    Achieving robust and scalable remote quantum entanglement is a fundamental challenge for the development of distributed quantum networks and modular quantum computing systems. Along this, perfect state transfer (PST) and fractional state transfer (FST) have emerged as promising schemes for quantum state transfer and remote entanglement generation using only nearest-neighbor couplings. However, the current implementations suffer from quantum loss and limited parameter tunability. In this work, we propose a new quantum state transfer scheme based on a zig-zag configuration, which introduces a controlling parameter for PST and FST. We show that this new parameter can suppress the population in the intermediate qubits, thereby reducing losses. We experimentally demonstrate the dynamics of different configurations on a superconducting quantum processor, achieving an 18\% reduction in error for remote Bell state generation in a 1D ($1\times5$) qubit chain, and exhibit robustness against certain types of noise. Then we extend our approach to a 2D network, successfully generating a W state among the four corner qubits. These results highlight the potential of our enhanced quantum state transfer scheme for scalable and noise-resilient quantum communication and computing.
\end{abstract}

\maketitle
\let\oldaddcontentsline\addcontentsline  
\renewcommand{\addcontentsline}[3]{}  

\section{Introduction}

Quantum computation and quantum simulation have entered the noisy intermediate-scale quantum (NISQ) era~\cite{2018_nisq_Preskill, 2019_nisq_Brooks}. Currently, quantum chips based on superconducting circuits~\cite{2025_newest_willow, 2019_newest_Sycamore, 2025_newest_Zuchongzhi3.1, 2025_newest_Zuchongzhi3.0, 2022_newest_Zuchongzhi2.1, 2021_newest_Zuchongzhi2.0, 2021_newest_Zuchongzhi1.0, 2025_newest_ZJU_11x11, 2025_newest_ZJU_2x20, 2024_newest_ZJU, 2022_newest_ZJU, 2025_newest_Chuang-tzu2.0, 2023_newest_Chuang-tzu, 2025_newest_FanHeng, 2024_newest_IBM, 2023_newest_IBM, 2022_newest_IBM, 2024_newest_Baqis, 2025_newest_YuDapeng, 2023_newest_YuDapeng}, photons~\cite{2021_newest_jiuzhang2.0, 2020_newest_jiuzhang1.0}, trapped ions~\cite{2023_newest_ions}, neutral atoms~\cite{2024_newest_atoms} and other platforms have integrated hundreds of qubits on a single chip, with several teams publishing roadmaps to realize chips with thousands of qubits in the coming years. 
Among these, superconducting quantum chips are advancing rapidly due to their ease of fabrication and scalability. However, the static nature of their qubits restricts interactions to short ranges, which hinders the efficient interconnection between distant qubits on the chip~\cite{2024_Wallraff_self-testing, 2023_Wallraff_Loophole-free, 2020_Wallraff_Quantum-Link, 2024_remote-gates_YuHaifeng, 2024_RIP_YuYang, 2024_plug-play_Michael, 2024_ZhongYoupeng_Anyonic, 2023_ZhongYoupeng_teleportation, 2023_ZhongYoupeng_Low-loss, 2021_ZhongYoupeng_deterministic, 2021_entangle-die_Gold, 2022_chiplet_Smith}. In this context, quantum state transfer and remote entanglement generation are essential for establishing remote qubit interconnections and, ultimately, enabling distributed quantum computing architectures~\cite{2024_Malekakhlagh_EnhancedPST}, which can be realized on superconducting quantum chips. 

Perfect state transfer (PST), in which an arbitrary state can be perfectly transferred from the leftmost site to the rightmost site, can be realized in a quantum circuit with only nearest-neighbor interactions~\cite{2004christandl, 2005wengwenkang, 2010kay}. Furthermore, fractional state transfer (FST), which can be regarded as an isospectral deformation of the PST Hamiltonian, enables the generation of remote entangled states between two distant sites \cite{2016genest}. With these ideas, the PST has been realized on various quantum hardware platforms~\cite{2018_LiX_PST, 2023_zhangchi_pst, 2024xiangliang, paritydependent_roy_2025, 2016_optical_Chapman, 2013_optical_perez, 2012_optical_bellec, 2005_NMR_ZhangJingfu, 2020_nanoelectromechanical_TianTian}.
However, due to dissipation induced by interaction with the environment, the implementation of quantum state transfer and remote entanglement generation based on these schemes faces significant challenges as the system scales to larger quantum networks.
    
To overcome these challenges, in this work, we propose a new quantum state transfer scheme based on a zig-zag configuration, which can reduce to the conventional PST scheme in certain limiting cases (referred to as the line configuration) while offering much broader applicability. We show that, compared to the line configuration, our scheme can suppress the state population in intermediate qubits, thereby reducing the influence of dissipation during the state transfer process and remote entanglement generation between distant sites. Specifically, using a 1D qubit chain, we achieve an 18\% reduction in error for remote Bell state generation compared to the traditional line configuration. Furthermore, we extend the application to a 2D ($3 \times 3$) qubit network, realizing the $W$ state among the four corner qubits, thus showcasing the scalability of our method. These results mark a significant step toward realizing the full potential of quantum state transfer and remote entanglement generation in next-generation quantum computing.
    
\section{Results}

We start from the following Hamiltonian with nearest-neighbor (NN) XY-type interaction (let $\hbar =1$)
\begin{equation}
    H = \sum_{n=1}^N\frac{\omega_n}{2}\sigma_n^z+\sum_{n=1}^{N-1}J_{n}(\sigma_n^+\sigma_{n+1}^-+\sigma_{n+1}^-\sigma_n^+)
    \label{eq-H},
\end{equation}
where $\sigma_n^{\pm,z}$ are the Pauli operators for qubit $Q_n$ with frequency $\omega_n$, and $J_{n}$ is the NN coupling between qubits $Q_n$ and $Q_{n+1}$. In superconducting qubits, as used in this work, the coupling strength can be adjusted using a tunable coupler~\cite{2018_YanFei_coupler, 2023_ZhangChi_coupler}. 

\begin{figure*}
    \centering
    \includegraphics[width=1.7\columnwidth]{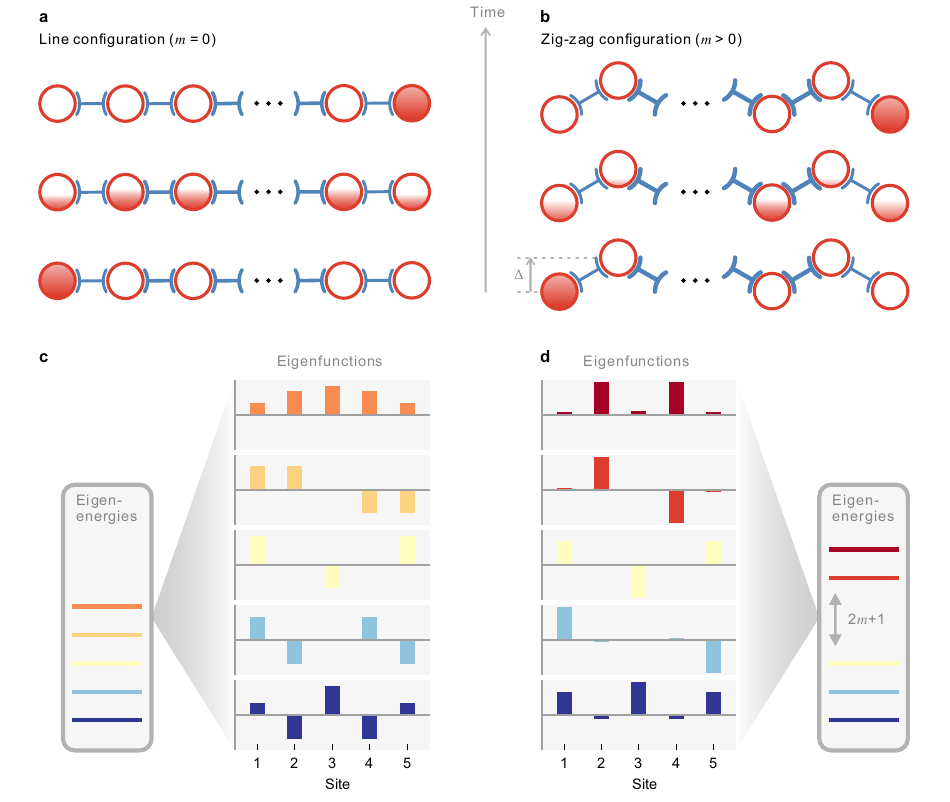}
    \caption{\textbf{Schematic representation.} \textbf{a, b} Conceptual diagram of the quantum state transfer. In the line configuration, all qubits share the same working frequency, and the population is distributed across all qubits during the evolution. In the zig-zag configuration, the working point frequencies feature an alternating pattern, with low frequencies for odd qubits and high frequencies for even qubits, leading to significant suppression of population on the even qubits during evolution.
        \textbf{c, d} Schematic of the eigenvalues and eigenfunctions, corresponding to the line configuration ($m=0$) and zig-zag configuration ($m>0$), respectively. For the line configuration, the eigenvalues are uniformly spaced with a spacing of 1. For the zig-zag configuration, an energy gap of $2m$ appears above the zero point in the positive half of the eigen-spectrum, and the eigenfunctions show significant amplitude suppression at specific indices.
    }
    \label{fig-fig1}
\end{figure*}

The goal of PST is to construct a Hamiltonian $H_{\text{PST}}$ by engineering the qubit frequencies $\{\omega_n\}$ and coupling strengths $\{J_n\}$ in Eq.~\eqref{eq-H}, such that a quantum state initialized at the starting point $\ket{1}$ can be perfectly transferred to the end point $\ket{N}$ after a proper time $\tau$. This expressed as $e^{-iH_{\text{PST}} \cdot \tau}\ket{1}=e^{i\phi}\ket{N}$. Here, we use the notation $\ket{n}$ to denote the single excitation state where only the $n$-th site is excited while all other sites remain in the ground state.  Ref.~\cite{2010kay} provides a necessary and sufficient condition for achieving PST with the following conditions
\begin{align}
    &\omega_n = \omega_{N+1-n}, \quad J_n^2 = J_{N-n}^2, \label{eq-mirror} \\
    &(\lambda_{n+1} - \lambda_{n})\tau = (2m_n+1)\pi, \quad m_n \in \mathbb{Z}^+,
    \label{eq-lambda}
\end{align}
where the eigenvalues $\{\lambda_n\}$ are ordered in increasing value, i.e., $\lambda_n<\lambda_{n+1}$.  It has been shown that when the coupling strengths $\{J_n\}$ are mirror symmetric about the central lattice site, and all qubit frequencies $\{\omega_n\}$ are resonant, for example~\cite{2004christandl}
\begin{equation}
    \omega_n = 0, \quad J_n = \frac{J}{2}\sqrt{n(N-n)}, 
    \label{eq-conventialPST}
\end{equation}
the conditions in Eq.~\eqref{eq-mirror} and Eq.~\eqref{eq-lambda} can be easily satisfied, yielding PST, which has been extensively investigated in experiments. 

The configuration for PST in Eq.~\eqref{eq-conventialPST}, which we refer to as the line configuration, lacks extra parameters that can mitigate dissipation effects. Thus, when applying this model to systems with realistic structures, dissipation can become significant and may completely ruin the PST. Here the major aim of this work is to realize PST using an extended configuration that include extra tunable parameters to combat the dissipation effect. In this configuration, eigenvalues are given by $\{ \lambda(m) \}=\{ -\frac{N-1}{2}, -\frac{N-1}{2} +1, -\frac{N-1}{2}+2, \cdots, -1, 0, 2m+1, 2m+2, \cdots, 2m+\frac{N-1}{2} \}$, where $N$ is odd and $m$ is a non-negative integer. The structure of this configuration is illustrated in Fig.~\ref{fig-fig1}a and \ref{fig-fig1}b. Through the inverse eigenvalue process~\cite{2006gladwell, 2005_iep_Chu}, we obtain 
\begin{align}
    \omega_n(m) &=\mu_n \cdot 2m \cdot J, \label{eq-wn}\\
    J_n(m) &=\frac{J}{2}\sqrt{[n + \mu_n2m][N-n + \mu_{n+1}2m]}, \label{eq-Jn}
\end{align}
where $\mu_n=1$ when $n$ is odd and 0 when $n$ is even. Clearly, Eq.~\eqref{eq-wn} and Eq.~\eqref{eq-Jn} satisfy the PST conditions in Eq.~\eqref{eq-mirror} and Eq. \eqref{eq-lambda}. Due to the alternating high and low frequencies of $\{\omega_n\}$, we refer to it as the zig-zag configuration. This configuration has two intriguing limits that help us understand its essential physics. Firstly, when $m=0$, this model is reduced to the line configuration. Secondly, when $m$ is large enough, the occupation of the even sites vanishes~\cite{hidden_hu_2024}, yielding
\begin{align}
\omega_{n}^{\text{eff}} &= -\frac{J(N-1)}{4}, \label{eq-limit_w}\\
J_{n}^{\text{eff}} &= -\frac{J}{2} \sqrt{\frac{n+1}{2}(\frac{N+1}{2}-\frac{n+1}{2})}, \label{eq-limit_J}
\end{align}where $n = 1, 3, 5, ..., N$. $J_{n}$ represents the coupling between the $n$-th and $(n+2)$-th sites. The detailed derivation is provided in Supplementary Section 5.
In this limit, the model reduces to the line configuration consisting of approximately half of the sites, $(N+1)/2$. Consequently, all configurations with different $m \in \mathbb{Z}^+$ can support PST and, under suitable conditions, enable the generation of remote entanglement. A detailed theoretical analysis of this configuration and its advantages will be discussed elsewhere~\cite{wtl_in_prep}.

\begin{figure*}
\centering
\includegraphics[width=1.7\columnwidth]{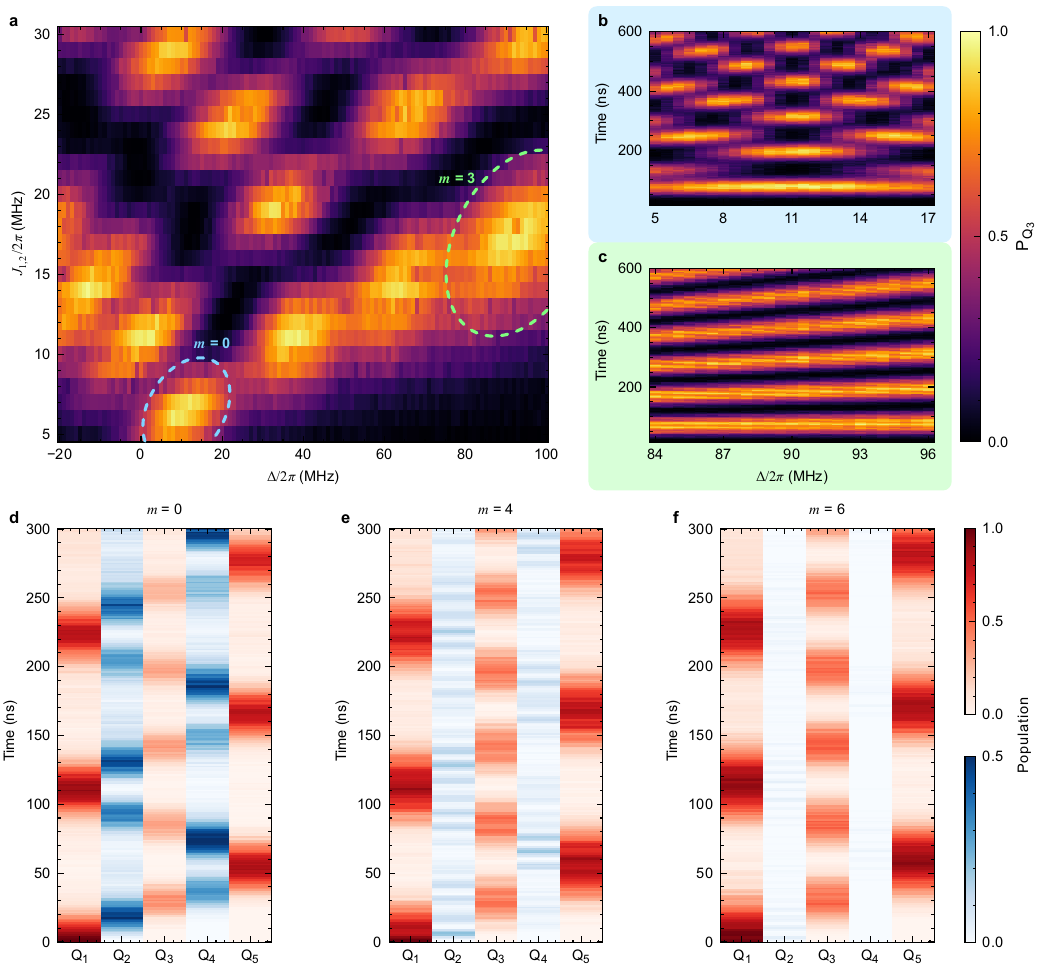}
    \caption{
        \textbf{Solution space and evolution spectra for quantum state transfer in 1D (1$\times$3 and 1$\times$5) qubit chains.}
        \textbf{a} Population spectrum for 1$\times$3 qubit state transfer, shown as a function of the coupling strength $J_{1,2}$ between qubits and the frequency detuning $\Delta$ between even and odd qubits. The colormap represents the population of $Q_3$, measured after preparing an excitation on $Q_1$ at the initial time and evolving for a fixed duration of 60~ns. Each bright spot represents a maximum population on $Q_3$, indicating the successful transfer of the excitation from $Q_1$ to $Q_3$. 
        \textbf{b-c} $Q_3$ population evolution as a function of $\Delta$, near the first (blue box) and fourth (green box) spots, respectively.
        \textbf{d-f} Quantum state transfer spectrum corresponding to $m=0, 4, 6$ on a 1$\times$5 qubit chain. One excitation is initialized on $Q_1$, and then the population evolution of all qubits is measured simultaneously. To enhance clarity, different scales and colormaps are used for odd qubits (red, with a maximum of 1.0) and even qubits (blue, with a maximum of 0.5). 
    }
    \label{fig-fig2}
\end{figure*}

We select the $1\times3$, $1\times5$ and $3\times3$ regions on the WuKong quantum chip for experimental demonstration. The detailed layout of this quantum chip and its corresponding parameters are presented in Supplementary Sections 1 and 2. We first validate the existence of PST in a $1\times 3$ qubit chain by selecting three qubits from the quantum chip, with the results shown in Fig.~\ref{fig-fig2}a. Initially, one excitation is prepared on the leftmost qubit $Q_1$, then we scan the coupling strengths $J_{1,2}$ (with $J_1 = J_2$ due to mirror symmetry) and the frequency detuning $\Delta$ of the even qubit $\omega_2$ relative to the odd ones $\omega_{1,3}$ ($\Delta=\omega_2-\omega_{1,3}$,  with $\omega_1 = \omega_3$ due to mirror symmetry).  After a fixed evolution time $\tau$=60~ns, we measure the population on the rightmost qubit $Q_3$, showing that the population of $Q_3$ reaches a maximum value (approximately unity) at the bright spots in Fig.~\ref{fig-fig2}a, indicating that the quantum state has been successfully transferred from $Q_1$ to $Q_3$, thereby the corresponding parameters represent a PST solution. This model can be solved analytically for a $1\times3$ qubit chain, yielding the population of $Q_3$ at time $\tau$ as
\begin{equation*}
P_{Q_3}= \frac{1}{4}(\cos\frac{\Omega \tau}{2} - \cos\frac{\Delta \tau}{2})^2 + \frac{1}{4}(\frac{\Delta}{\Omega} \sin\frac{\Omega \tau}{2} - \sin\frac{\Delta\tau}{2})^2,
\end{equation*}
where $\Omega^2 = \Delta^2 + 8J_{1,2}^2$. Clearly, PST is achieved when $P_{Q_3}=1$, which occurs when the above condition is satisfied; see Eq.~\eqref{eq-wn} and Eq.~\eqref{eq-Jn}. The derivation and numerical results are presented in Supplementary Section 6. 

The first four bright spots from left to right in the first row correspond one-to-one to the analytical solutions from $m=0$ to $m=3$ given by Eq.~\eqref{eq-wn} and Eq.~\eqref{eq-Jn}. Note that the first spot should, in principle, occur at exactly zero detuning. However, in this experiment, the frequency shift induced by the couplers was not compensated, resulting in the first spot occurring at approximately 10~MHz. From these experimental results, we extract the coupling strength corresponding to the first bright spot as 
$\tilde{J}_{1,2}(0) / 2\pi \approx 6$~MHz. This value can also be obtained analytically via $J/2\pi = 1/2\tau = 8.3$~MHz, which yields $J_{1,2}(0) / 2\pi = \frac{J}{2} \sqrt{2} = 5.9$~MHz. Thus, the above two estimations are consistent with each other. Similarly, we calculate the value of $J_n(m)$ for $m = 1, 2, 3$, yielding $J_{1,2}(1) / 2\pi = 10.2$~MHz, $J_{1,2}(2) / 2\pi = 13.2$~MHz and $J_{1,2}(3) / 2\pi = 15.6$~MHz, all of which are in good agreement with the experimental results. 

In Fig.~\ref{fig-fig2}b and \ref{fig-fig2}c, we perform a fine scan of the state transfer population as a function of detuning $\Delta$ and duration time $\tau$ near the solutions for $m=0$ and $m=3$, respectively. For $m = 0$, population in the second site is allowed, yielding the fragmented structure in Fig. \ref{fig-fig2}b. However, when $m$ becomes large, such structure is suppressed, leading to the continuous pattern shown in Fig.~\ref{fig-fig2}c. We further experimentally compare the performance of PST in a $1\times 5$ qubit chain under different solutions for $m$. Fig.~\ref{fig-fig2}d--~\ref{fig-fig2}f show the dynamical evolution of each qubit for $m=0,4,6$, respectively. As $m$ increases, a noticeable reduction in the population of the even qubits is observed. For $m=6$, almost no visible population remains on these qubits, indicating that the noise channels of the even qubits have been effectively shut off. 

\begin{figure*}             
    \centering
    \includegraphics[width=1.96\columnwidth]{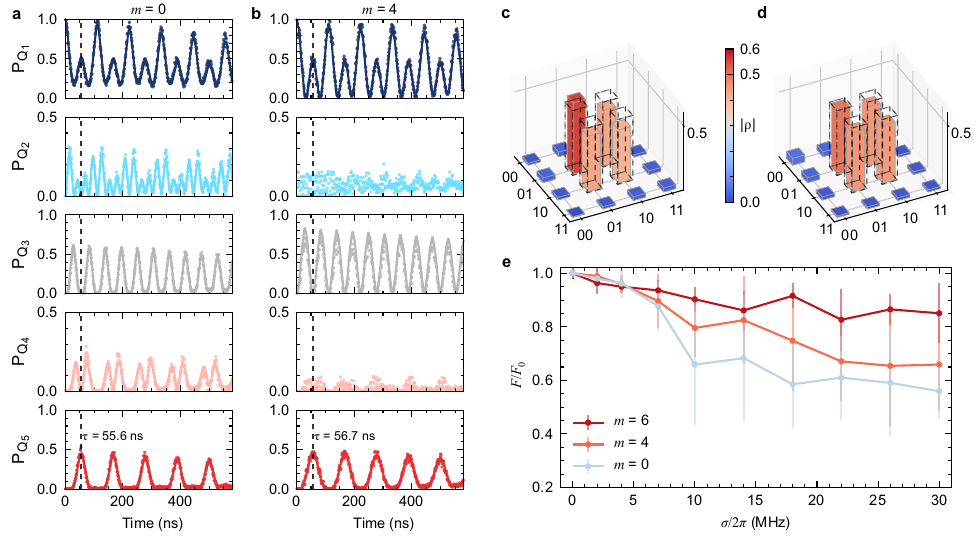}
    \caption{
        \textbf{Fractional state transfer and remote entanglement generation at various $\bm{m}$ values in a 1D (1$\times$5) qubit chain.}
        \textbf{a (b)} Fractional state transfer for $m=0$ ($m=4$). Initially, one excitation is prepared on $Q_1$, and at time $t$, the population of all qubits is measured simultaneously. The target transfer time is $\tau=55.6$~ns, though deviations may occur in practice after the parameter optimization process. Dots represent the experimental data, while the solid lines correspond to the master equation simulations.
        \textbf{c (d)} Quantum state tomography at $\tau$=55.6~ns ($\tau$=56.7~ns) for $m=0$ ($m=4$). The reconstructed density matrix yields the Bell state fidelity of 0.909$\pm$0.024 (0.925$\pm$0.021). The ideal Bell singlet state is $\ket{\Psi^-} = (\ket{01} - \ket{10}) / \sqrt{2}$, represented by the dashed box. The standard deviation in fidelity is obtained from 50 experimental repetitions.
        \textbf{e} Fidelity degradation, $F/F_0$ (with $F_0$ being the fidelity at zero noise), as a function of frequency noise applied to even qubits. The noise follows a Gaussian distribution $N(0, \sigma)$, where $\sigma$ is the noise intensity centered on the respective frequencies of the even qubits. 
    }
    \label{fig-fig3}
\end{figure*}

Based on the above results, which validate our theoretical proposal, we next demonstrate the remote entanglement generation using the zig-zag configuration and investigate the dissipation effect for various $m$. We follow the idea of fractional state transfer (FST), proposed by Genest et al.,~\cite{2016genest} i.e., $e^{-iH_{\text{FST}} \cdot \tau}\ket{1}=e^{i\phi}(\sin{2\theta}\ket{1} + \cos{2\theta}\ket{N})$. When $\theta = \pi/8$, the system evolves into a maximally entangled state between the remote sites $\ket{1}$ and $\ket{N}$. The PST model in Eq.~\eqref{eq-wn} and Eq.~\eqref{eq-Jn} cannot be directly used for this purpose, requiring the following isospectral transformation 
\begin{align}
    \tilde{J}_{\frac{N-1}{2}}(m) &= (\cos{\theta} + \sin{\theta}) J_{\frac{N-1}{2}}(m), \nonumber \\
    \tilde{J}_{\frac{N+1}{2}}(m) &= (\cos{\theta} - \sin{\theta}) J_{\frac{N+1}{2}}(m), \label{eq-Jfst}
\end{align}
while the frequencies and all the other coupling terms remain the same as Eq.~\eqref{eq-wn} and Eq.~\eqref{eq-Jn}.

Experimentally, we select $N=5$ and $\theta=\pi/8$ and engineer the Hamiltonian according to Eq.~\eqref{eq-wn}, Eq.~\eqref{eq-Jn} and Eq.~\eqref{eq-Jfst}. The periodic FST dynamics are shown in Fig.~\ref{fig-fig3}a and Fig.~\ref{fig-fig3}b, corresponding to $m=0$ and $m=4$, respectively. Initially, one excitation is initialized on $Q_1$, then the quantum state is fractionally transferred from $Q_1$ to both $Q_1$ and $Q_5$, after which it gradually transfers back to $Q_1$, completing an FST cycle. For $m=4$, the population of the even qubits is significantly suppressed compared to $m=0$, thereby protecting the coherence of the system. At $\tau$=55.6~ns for $m=0$ (56.7~ns for $m=4$), the population of the remote qubits, $Q_1$ and $Q_5$, approaches 0.5, while the population of the intermediate qubits is close to 0, suggesting that maximal entanglement has been achieved between $Q_1$ and $Q_5$. To further confirm the result, we perform the quantum state tomography (QST)~\cite{qst_2012_Smolin} on $Q_1$ and $Q_5$ in Fig.~\ref{fig-fig3}c, confirming that they are indeed in the Bell singlet state $\ket{\Psi^-} = (\ket{01} - \ket{10}) / \sqrt{2}$~\cite{2024xiangliang, 2020ChangHS}. Results in Fig.~\ref{fig-fig3}c and \ref{fig-fig3}d reveal that the QST fidelity for $m=0$ is 0.909$\pm$0.024, while for $m=4$, it is higher at 0.925$\pm$0.021. All experiments with different $m$ values are carried out on the same set of qubits within the same time period, demonstrating that increasing $m$ can suppress the dissipation effect, which is essential for PST and FST for longer chains. 

In Fig.~\ref{fig-fig3}e, we further confirm the advantage of increasing $m$ for noise suppression. To this end, we intentionally apply Gaussian noise $N(0, \sigma)$ with intensity $\sigma$ centered on the respective frequencies of even qubits~\cite{2024xiangliang, drift_2018_MeiFeng}, and then measure the QST fidelity degradation (i.e., $F/F_0$, where $F_0$ is the QST fidelity at zero noise) as a function of $\sigma$ for different $m$. As the noise intensity increases from 0 to 30~MHz, the fidelity for $m=0$ drops rapidly. In contrast, for $m=4$ and $m=6$, the fidelities degrade much more slowly. For instance, when $\sigma/2\pi = 30$~MHz, these values for $m = 0$, 4, 6 are 0.56, 0.66 and 0.85, respectively. This is because as $m$ increases, the even sites are much less occupied, and when $m$ are large enough, the effective chain is reduced by half, as shown in Eq.~\eqref{eq-limit_w} and Eq.~\eqref{eq-limit_J}. For detailed experimental data and corresponding simulation results on the robustness of the zig-zag configuration against other types of noise, refer to Supplementary Section 9.

\begin{figure}
    \centering
    \includegraphics[width=\columnwidth]{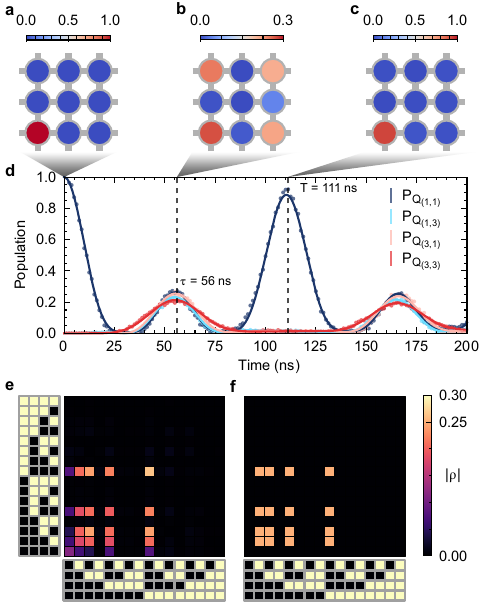}
    \caption{
        \textbf{Fractional state transfer and remote entanglement generation in a 2D (3$\times$3) qubit network.}
        \textbf{d} Fractional state transfer. The population of all qubits are measured simultaneously at different times, with $Q_1$ initially prepared in a single excitation state. For clarity, only the population of the four corner qubits are plotted. The dots represent the experimental data, while the solid lines indicate the results from master equation simulations. 
        \textbf{a--c} Population maps of the $3\times3$ qubit network measured at 0, 56~ns and 111~ns, respectively.
        \textbf{e} Reconstructed density matrix of the four corner qubits from the quantum state tomography experiment. The left and bottom axes show matrix indices represented by binary strings, where black denotes 0 and light yellow denotes 1. The indices increase from 0000 to 1111 from left to right and bottom to top.  
        \textbf{f} The ideal density matrix of 4-qubit W state, yielding a state fidelity of 0.85.}
    \label{fig-fig4}
\end{figure}

Finally, we demonstrate the scalability of FST by extending our experiment to a 2D qubit network. This extension is achieved by generalizing the 1D model to 2D with parameters \{$\omega_n^{x,y}$\} and \{$J_n^{x,y}$\}~\cite{2010kay}, where the FST condition Eq.~\eqref{eq-Jfst} is satisfied along each dimension. Fig.~\ref{fig-fig4}d shows the population evolution of the four corner qubits over time by initially preparing an excitation on $Q_1$ at the lower-left corner, while the Fig.~\ref{fig-fig4}a--~\ref{fig-fig4}c display the population distribution of all $3\times3$ qubits at 0, 56~ns and 111~ns, respectively. At 56~ns, the excitation is uniformly transferred to the four corner qubits, suggesting that the system has evolved into a maximally entangled state between these qubits. By 111~ns, the system returns to the single excitation state on $Q_1$, and this fractional revival process repeats periodically in subsequent evolution, providing experimental validation of the effectiveness of FST theory in the 2D case. We further perform quantum state tomography (QST) at 56~ns to reconstruct the density matrix, as shown in Fig.~\ref{fig-fig4}e, while Fig.~\ref{fig-fig4}f shows the theoretical density matrix $\rho = \ket{\mathrm{W}}\bra{\mathrm{W}}$, corresponding to the maximally entangled W state as~\cite{2010_w_neeley}
\begin{equation}
\ket{\mathrm{W}} = {1\over 2}(\ket{1000} + \ket{0100} + \ket{0010} + \ket{0001}).
\label{eq-Wstate}
\end{equation}
The experimentally reconstructed density matrix shows excellent agreement with the theoretical value. Decoherence effects introduce unexpected color blocks, thereby reducing the QST fidelity. We find that the measured fidelity is 0.85.

\section{Discussion and Conclusion}
To conclude, in this work we propose a novel scheme for perfect state transfer (PST) and fractional state transfer (FST) based on a zig-zag configuration, enabling quantum state transfer and remote entanglement generation. As compared with the previous approach, this scheme offers a new degree of freedom -- the energy difference between the even and odd sites, controlled by parameter $m$ -- which can effectively suppress the channel noise by reducing the occupation of the intermediate sites. A key feature of this model is that, when the energy difference between odd and even sites become sufficiently large, it will reduce to the conventional PST model by effectively integrating out the even sites. Using this model, we demonstrate two key applications on systems with 3, 5 and 9 qubits. In the first application, we experimentally demonstrate the PST and show that the energy difference associated with $m$ can help suppress the dissipation. In the second application, we demonstrate the remote entanglement via FST in both a 1D qubit chain and 2D qubit network, generating a two-qubit Bell state and a four-qubit W state, respectively. 

Our scheme can be generalized to much larger quantum systems \cite{2024xiangliang}, and adapted to other quantum platforms. Furthermore, we anticipate the development of more versatile Hamiltonians tailored to specific experimental needs~\cite{2023_Hengineering_Inui, 2024_Hengineering_Inui}, such as applications in logical state preparation, quantum simulation and other research areas. These realizations are expected to be important for future quantum computers, which may soon scale to thousands of qubits. In such large-scale systems, remote state transfer and remote entanglement generation will become essential experimental techniques.

\begin{acknowledgments}
This work is supported by the Strategic Priority Research Program of the Chinese Academy of Sciences (Grant No.~XDB0500000), the National Natural Science Foundation of China (Grant No.~12404564 and No.~U23A2074) and the Innovation Program for Quantum Science and Technology (2021ZD0301200, 2021ZD0301500). This work is also partially carried out at the USTC Center for Micro and Nanoscale Research and Fabrication. 
\end{acknowledgments}

\bibliography{ref.bib}

\clearpage
\beginsupplement
\let\addcontentsline\oldaddcontentsline

\onecolumngrid
\begin{center}
    \textbf{\large Supplementary materials for "Remote entanglement generation via enhanced quantum state transfer"}
\end{center}
\vspace{0.5cm}

\tableofcontents

\section{Device informatioin}
    Our experiments are conducted on the ``Wukong'' quantum processor, a two-dimensional (2D) flip-chip superconducting quantum chip consisting of 72 qubits and 126 couplers arranged in a $6\times12$ rectangular grid. All qubits and couplers are designed with asymmetric SQUID structures. Each qubit is equipped with an independent XY control line for driving single qubit rotation, an independent Z control line for tuning its frequency, and an individual readout resonator for measuring the qubit state. Each coupler, on the other hand, has only one single Z control line for adjusting its frequency and thereby varying the coupling strength between the two connected qubits. 
    
    For the 1D and 2D perfect state transfer (PST) and fractional state transfer (FST) experiments, a total of 13 qubits are utilized. The idle frequencies of all 65 available qubits, including those used in our experiments, are allocated using a snake-type optimizer~\cite{2020_snake_Klimov}. The labels of these qubits and their positions on the chip are shown in Fig.~\ref{fig-figs1}. The performance of these qubits, including idle/max/min frequencies $\omega_{\text{idle/max/min}}$, energy relaxation time $T_1$, Ramsey dephasing time $T_2^{\ast}$, and readout fidelities for the 0/1 states $F_{0/1}$, are summarized in Table.~\ref{tab:device}
    
\section{Wiring}
\begin{figure}
        \centering
        \includegraphics[width=0.9\columnwidth]{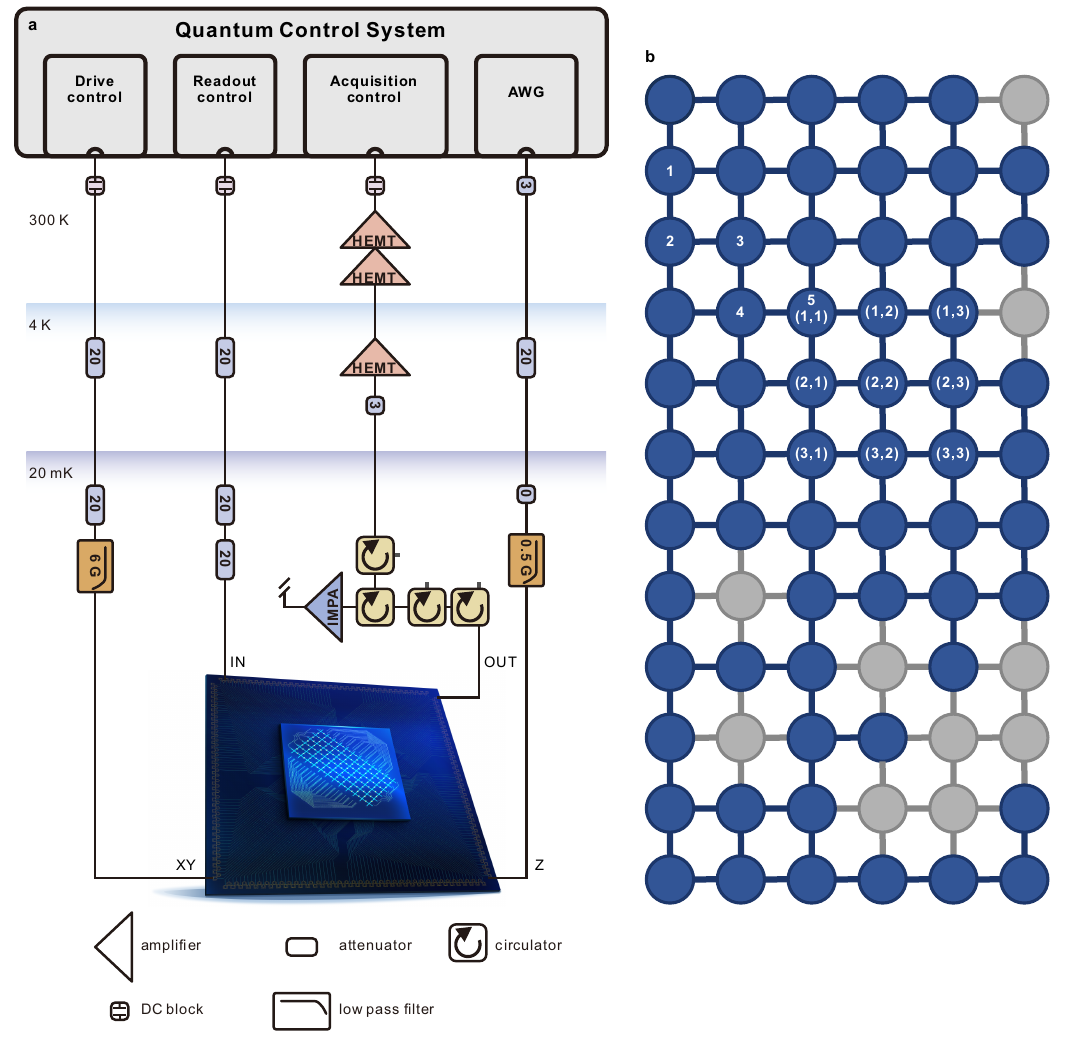}
        \caption{
            \textbf{Schematic of the cryogenic wiring and chip layout. }
            \textbf{a} Control signals are emitted from room-temperature quantum control system and pass through a set of attenuators, filters and other components before reaching the quantum chip. Signals returning from the chip are amplified by an impedance-transformed parametric amplifier (IMPA), a cryogenic high electron mobility transistor (HEMT), and two room-temperature HEMTs before being received by the quantum control system. Note that the IMPA requires a DC flux provided by a voltage source and a pump tone from a microwave source, which are combined via a room-temperature bias-tee before being fed to the IMPA. 
            \textbf{b} The ``Wukong'' quantum chip features a $12\times6$ architecture, in which each readout resonator is shared by six qubits. Nonfunctional qubits and couplers are marked in gray. 
        }
        \label{fig-figs1}
    \end{figure}

    The quantum processor is housed in a dilution refrigerator, with the mixing chamber (MC) plate maintained at approximately 20~mK. The processor is connected to room-temperature control and measurement electronics through multiple stages of attenuators and filters for noise suppression. These electronics are integrated into a quantum control system. The arbitrary waveform generator (AWG) module, with a sampling rate of 1.2~GSa/s, delivers both static flux bias and fast flux pulses to the qubits. The radio frequency (RF) module, with a sampling rate of 3.2~GSa/s, can output signals ranging from 3.75 to 8.25~GHz to drive the qubits and generate readout pulses. For readout acquisition, every six qubits share one readout transmission line. The readout signal is first amplified by an impedance-transformed parametric amplifier (IMPA)~\cite{impa_duan_2021} at the MC state, then by a high electron mobility transistor (HEMT) at 4~K stage, and finally by room-temperature amplifiers before entering the acquisition module. After IQ demodulation, the signal is used to distinguish the quantum states. Detailed wiring configuration is illustrated in Fig.~\ref{fig-figs1}.

    \begin{table*}[t] 
        \renewcommand\tabcolsep{12.0pt}
        \begin{tabular}{c|cccccccc}
        \hline
        \hline
        Qubit & $\omega_{\mathrm{idle}}/2\pi$ & $\omega_{\mathrm{max}}/2\pi$ & $\omega_{\mathrm{min}}/2\pi$ & $T_1$ & $T_2^{\ast}$ & $F_0$ & $F_1$ \\
        Index & (GHz) & (GHz) & (GHz) & ($\mu$s)& ($\mu$s) &  &  \\
        \hline
        $Q_1$           & 4.095 & 4.317 & 3.861 & 7.2  & 1.2 & 0.930 & 0.822 \\
        $Q_2$           & 4.658 & 4.658 & 4.123 & 13.4 & 4.1 & 0.965 & 0.763 \\
        $Q_3$           & 4.000 & 4.367 & 3.901 & 13.9 & 0.7 & 0.929 & 0.782 \\
        $Q_4$           & 4.143 & 4.606 & 4.092 & 16.6 & 1.6 & 0.923 & 0.821 \\
        $Q_5/Q_{(1,1)}$ & 4.122 & 4.373 & 3.898 & 29.4 & 1.0 & 0.934 & 0.834 \\
        $Q_{(1,2)}$     & 4.653 & 4.729 & 4.166 & 13.2 & 1.3 & 0.949 & 0.801 \\
        $Q_{(1,3)}$     & 4.106 & 4.297 & 3.796 & 11.0 & 1.6 & 0.925 & 0.819 \\
        $Q_{(2,1)}$     & 4.240 & 4.603 & 4.097 & 13.9 & 2.0 & 0.951 & 0.857 \\
        $Q_{(2,2)}$     & 4.130 & 4.432 & 3.983 & 14.0 & 6.8 & 0.920 & 0.857 \\
        $Q_{(2,3)}$     & 4.690 & 4.780 & 4.236 & 7.3  & 2.0 & 0.954 & 0.863 \\
        $Q_{(3,1)}$     & 4.020 & 4.360 & 3.882 & 13.6 & 1.6 & 0.922 & 0.837 \\
        $Q_{(3,2)}$     & 4.460 & 4.624 & 4.123 & 11.0 & 1.3 & 0.910 & 0.850 \\
        $Q_{(3,3)}$     & 4.190 & 4.428 & 3.935 & 13.9 & 1.7 & 0.924 & 0.846 \\
        \hline
        \hline
        \end{tabular}
        
        \caption{
            \textbf{Qubit parameters.} $\omega_{\text{idle}}, \omega_{\text{max}}, \omega_{\text{min}}$ denote the qubit frequencies at the idle, maximum and minimum point, respectively. $T_1$ and $T_2^\ast$ represent the relaxation time and Ramsey dephasing time at the idle point, respectively. $F_0$ and $F_1$ denote the readout fidelities of the qubit states $\ket{0}$ and $\ket{1}$, respectively. 
        } 
        \label{tab:device}
    \end{table*}

\section{Theory of perfect state transfer}

    The perfect state transfer (PST) discussed in the main text refers to the process whereby a quantum state, under the evolution of a pre-engineered Hamiltonian, can be transferred perfectly from one end (or one state) of a quantum chip to the other end (or other state). Specifically, after an evolution time $\tau$, we have 
    \begin{align}
        e^{-iH_{\text{PST}} \cdot \tau}\ket{1}=e^{i\phi}\ket{N},
    \end{align}
    where $\ket{n}$ denotes a single excitation at the $n$-th qubit ($n=1,2,\cdots,N$), and $\phi$ is a global phase irrelevant to the measurement. 

    In 2004, Christandl et al.~\cite{2004christandl} proposed a theory for PST that corresponds to the line configuration discussed in the main text. They showed that PST can be achieved in a 1D qubit chain, when the single-excitation subspace of the Hamiltonian is equivalent to a $(N-1)/2$-spin operator. In 2010, Kay~\cite{2010kay} further summarized the necessary and sufficient conditions for PST , which can be formulated as the mirror symmetry condition
    \begin{equation}
        \omega_n = \omega_{N+1-n}, \quad J_n^2 = J_{N-n}^2, \label{eq:mirror}
    \end{equation}
    and the eigenvalue spacing condition
    \begin{equation}
        (\lambda_{n+1} - \lambda_{n})\tau = (2m_n+1)\pi,  \quad m_n \in \mathbb{Z} \label{eq:lambda}
    \end{equation}
    where $\{\omega_n\}$ and $\{J_n\}$ represent the frequency of the $n$-th qubit and the coupling strength between the $n$-th and $(n+1)$-th qubits, respectively. The eigenvalues of the Hamiltonian, denoted by $\{\lambda_n\}$, are arranged in ascending order of energy.
    These conditions imply that PST can be realized, as long as the designed Hamiltonian and its eigenvalues satisfy the above two conditions.

\section{Theory of fractional state transfer \label{sec:FST}}

    Genest et al. proposed the theory of fractional state transfer (FST) in 2016~\cite{2016genest}, which enables an excitation initialized at one site to evolve fractionally to both itself and its mirror site. This mechanism provides a powerful tool for generating remote entangled states in a continuous and tunable manner. Here, we provide a brief summary of the FST theory used in the main text.

    Given the initial state $\ket{1}$, the evolution of the FST Hamiltonian requires \cite{2016genest}
    \begin{align}
        e^{-iH_{\text{FST}} \cdot \tau}\ket{1}=e^{i\phi}(\sin{2\theta}\ket{1} + \cos{2\theta}\ket{N}), \label{eq:fst}
    \end{align}
   in which the FST Hamiltonian share the same eigenvalues as PST. Therefore, we seek to identify a unitary transformation $U$ such that $H_{\text{FST}} = U H_{\text{PST}} U^{\dagger}$. The time evolution then satisfies
    \begin{align*}
        e^{-iH_{\text{FST}} \cdot \tau} = U e^{-iH_{\text{PST}} \cdot \tau} U^{\dagger} = e^{i \phi} URU^{\dagger} = e^{i \phi} Q,
    \end{align*}
    where $R$ is both the symmetry operator and the evolution operator for PST after eliminating the global phase, with the form $R = \sum_{n=1}^{N} \kbra{n}{N+1-n}$. Q is the evolution operator for FST (excluding the global phase), with its first column given by Eq.~\eqref{eq:fst}. The remaining columns can be readily obtained by applying the evolution to other basis states. For mathematical simplicity, we write $U$ as $U=VR$, the explicit form of $V$ is given by 
    \begin{equation*}
        V = \begin{pmatrix}
          \sin \theta &  &  &  &  &  & \cos \theta \\
          & \ddots &  &  &  & \iddots & \\
          &  & \sin \theta & 0 & \cos \theta &  & \\
          &  & 0 & 1 & 0 &  & \\
          &  & \cos \theta & 0 & -\sin \theta &  & \\
          & \iddots &  &  &  & \ddots  & \\
          \cos \theta &  &  &  &  &  & -\sin \theta
        \end{pmatrix},
    \end{equation*}
    when $N$ is odd, and     
\begin{equation*}
        V = \begin{pmatrix}
          \sin \theta &  &  &  &  & \cos \theta \\
          & \ddots &  &  & \iddots & \\
          &  & \sin \theta & \cos \theta &  & \\
          &  & \cos \theta & -\sin \theta &  & \\
          & \iddots &  &  & \ddots  & \\
          \cos \theta &  &  &  &  & -\sin \theta
        \end{pmatrix},
\end{equation*}
when $N$ is even. It is straightforward to verify that $V$ is both Hermitian and unitary, i.e., $V=V^{\dagger}$ and $VV^{\dagger}=I$. Furthermore, one can verify that $Q=VRV$, which implies that $QQ^{\dagger}=I$. In fact, the matrix elements of $Q$ can be conveniently obtained by replacing $\theta$ in $V$ with $2\theta$, clearly reflecting the evolution characteristics of FST. 

    Therefore, once the form of $H_{\text{PST}}$ is determined, $H_{\text{FST}}$ can be derived through a simple isospectral deformation. Examing the form of $H_{\text{FST}}$ reveals that, compared to $H_{\text{PST}}$, only a few modifications are required in the middle terms
    \begin{align}
        &\left\{\begin{matrix}
        &\tilde{J}_{\frac{N-1}{2}} &=& (\cos{\theta} + \sin{\theta}) J_{\frac{N-1}{2}} \\
        &\tilde{J}_{\frac{N+1}{2}} &=& (\cos{\theta} - \sin{\theta}) J_{\frac{N+1}{2}}
        \end{matrix}\right., \label{eq:Hfst_odd}
    \end{align}
    when $N$ is odd, and 
    \begin{align}
        &\left\{\begin{matrix}
        &\tilde{J}_{\frac{N}{2}} &=& \cos{2\theta} J_{\frac{N}{2}} \\
        &\tilde{\omega}_{\frac{N+1\pm1}{2}} &=& \omega_{\frac{N}{2}} \pm \sin{2\theta} J_{\frac{N}{2}}
        \end{matrix}\right., \label{eq:Hfst_even}
    \end{align}
    when $N$ is even, where the tilde denotes the parameters of $H_{\text{FST}}$. Regardless of the system size $N$, only a few specific parameters -- those shown in Eq.~\eqref{eq:Hfst_odd} and \eqref{eq:Hfst_even} -- are changed when transitioning from $H_{\text{PST}}$ to $H_{\text{FST}}$. This significantly simplifies the parameter calibration process in experiments.

\section{The zig-zag configuration}
    In this work, we introduce a new model with zig-zag configuration to realize the FST and PST. According to the inverse eigenvalue theory~\cite{2006gladwell, 2005_iep_Chu}, once a specific set of eigenvalues is provided, a unique mirror-symmetric tridiagonal Hamiltonian can be reconstructed. We therefore heuristically propose the following eigenvalues 
    \begin{align*}
        \{ \lambda(m) \}= -\frac{N-1}{2}, \cdots, -1, 0, 2m+1, \cdots, 2m+\frac{N-1}{2},
    \end{align*}
    in which the underlying design philosophy is to introduce an energy gap that separates different subspaces, thereby suppressing population on certain energy levels during the evolution. Moreover, with this design, both the eigenvalue spacing condition and the mirror symmetry condition are naturally fulfilled. 

    The method of reconstructing the Hamiltonian via eigenvalues has been comprehensively summarized in Ref.~\cite{2006gladwell} and Ref.~\cite{2005_iep_Chu}. We will also provide a detailed description of this method in a forthcoming publication~\cite{wtl_in_prep}. For a qubit chain of length $N$, starting from $n=1$, we can iteratively construct the target frequency terms $\omega_n$ and coupling strength terms $J_n$. By repeating this process over different values of length $N$, we can generalize the analytical expressions for $\omega_n$ and $J_n$ as functions of arbitrary $N$ and $n$, namely
    \begin{align}
        \omega_n(m) &=\mu_n \cdot 2m \cdot J, \label{eq:wn}\\
        J_n(m) &=\frac{J}{2}\sqrt{[n + \mu_n2m][N-n + \mu_{n+1}2m]}, \label{eq:Jn}
    \end{align}
    where $\mu_n=1$ when $n$ is odd and 0 when even.
    
    The Hamiltonian reconstructed above is exclusively capable of realizing PST. Moreover, as demonstrated in Sec.~\ref{sec:FST}, an isospectral deformation of a PST Hamiltonian can yield a fractional state transfer (FST) Hamiltonian, which enables the generation of remote entanglement. Specifically, only the coupling terms at the middle sites need to be modified, as shown in Eq.~\eqref{eq:Hfst_odd} and Eq.~\eqref{eq:Hfst_even}.

    We now discuss the asymptotic properties of the zig-zag configuration. When $m=0$, the Hamiltonian reduces naturally to the line configuration. For large values of $m$, we can integrate the even sites using the method in Ref.~\cite{hidden_hu_2024}. For the site 1, 2 and 3, we have 
    \begin{equation}
    J_1 c_1 + J_2 c_3 + \omega_2 c_2 = E c_2, \quad c_2 = {J_1 c_1 + J_2 c_3 \over E-\omega_2}.
    \end{equation}
    
    For site 3, 4, 5, and im a similar way, we have 
    \begin{equation}
    J_3 c_3 + J_4 c_5 + \omega_4 c_4 = E c_4, \quad c_4 = {J_3 c_3 + J_4 c_5 \over E-\omega_4}.
    \end{equation}
    
    Then we can obtain the equation of motion for sites 2, 3, 4, using 
    \begin{equation}
    J_2 c_3 + J_3 c_4 + \omega_3 c_3 = E c_4,
    \end{equation}
    and substituting $c_2$, $c_4$ into the above model will yields a new eigenvector equation involving only sites 1, 3, 5, as following
    \begin{equation*}
        \frac{J_1 J_2}{E - \omega_2}c_1 + (\frac{J_2^2}{E - \omega_2} + \frac{J_3^2}{E - \omega_4}) c_3 + \frac{J_3 J_4}{E - \omega_4} c_5 = E c_3.
    \end{equation*}
    
    This idea can be generalized to the whole chain by writing an effective tight-binding model for only the odd sites, yielding the effective coupling strengths and frequency corrections as 
    \begin{subequations}
    \begin{align}
        J_{n}^{\text{eff}} &= \frac{J^2}{4} \frac{\sqrt{(n + 2m)(N-n)} \sqrt{(n+1)(N-n-1 + 2m)}}{E - 2mJ},\\
        \omega_n^{\text{eff}} &= \frac{J^2}{4}\left[\frac{(n-1)(N-n+1 + 2m)}{E - 2mJ} + \frac{(n + 2m)(N-n)}{E - 2mJ}\right],
    \end{align}
    \end{subequations}
    in which we have defined $n=2n'-1$, with $n'=1,2,\cdots,(N+1)/2$. 

    When $m$ is large enough, let $m\to\infty$, we can neglect $E$ in the denominator and have 
    \begin{subequations}
    \begin{align}
        \left.J_{n}^{\text{eff}}\right\vert_{m\to\infty} &= -\frac{J}{4} \sqrt{(n+1)(N-n)} 
        = -\frac{J}{2}\sqrt{n'(\frac{N+1}{2}-n')},\\
        \left.\omega_{n}^{\text{eff}}\right\vert_{m\to\infty} &= -\frac{J}{4}(N-1),
    \end{align}
    \end{subequations}
    This is just the PST proposed by Christandl et al.~\cite{2004christandl} for a chain with length $(N+1)/2$. 
    
\section{Simulation of the PST solution space}
    \begin{figure}
        \centering
        \includegraphics[width=0.8\columnwidth]{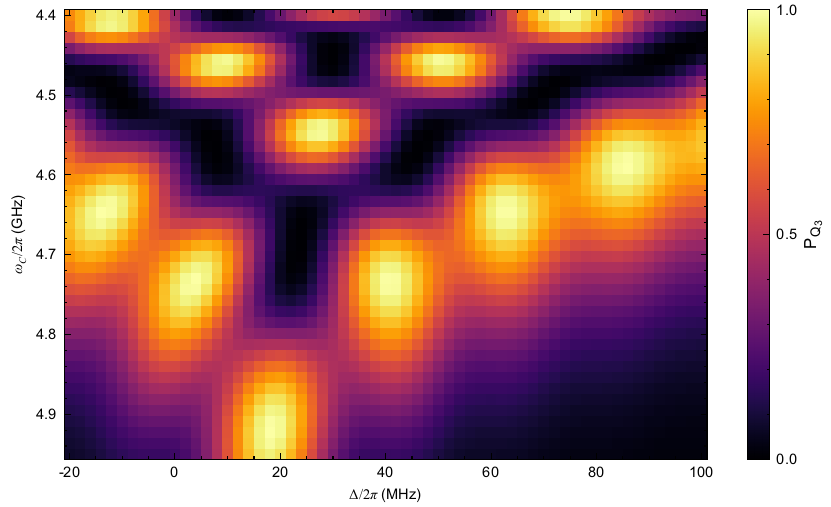}
        \caption{
            \textbf{Simulation of the solution space for quantum state transfer in a $1\times3$ qubit chain.} The population of the end qubit $Q_3$ is plotted as a function of the frequency of $Q_2$ ($\omega_{Q_2}$) and the coupler frequencies ($\omega_{C}$), which determine the coupling strengths between qubits. The evolution time is fixed at 60~ns. The simulation result shows a good agreement with the experimental data presented in Fig.~2a of the main text. Each bright spot indicates a successful transfer from $Q_1$ to $Q_3$. The parameters corresponding to the first row of bright spots match the solutions for $m=0$ to $m=3$ given in the main text.
        }
        \label{fig-figs2}
    \end{figure}

    \begin{figure}
        \centering
        \includegraphics[width=0.98\columnwidth]{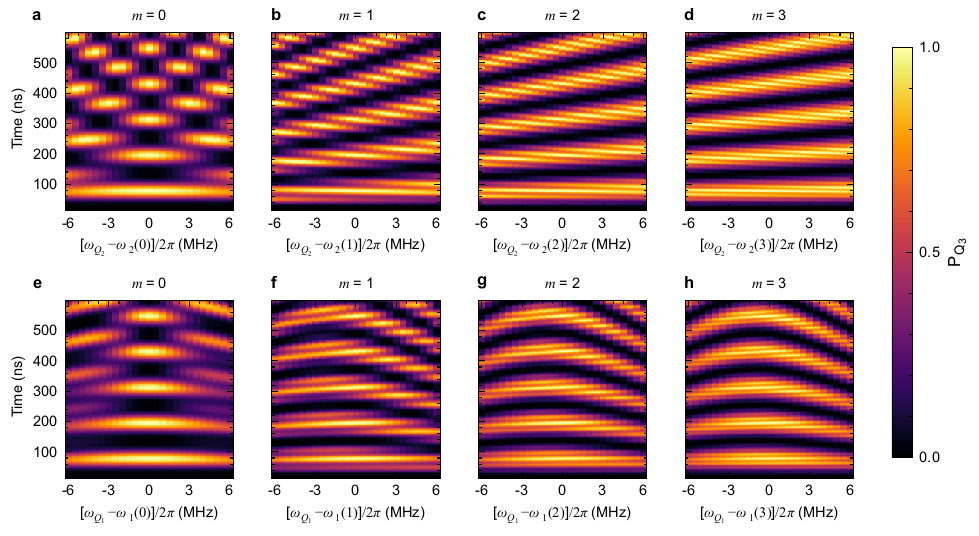}
        \caption{
            \textbf{Simulation of the evolution spectrum for quantum state transfer in a 1$\times$3 qubit chain.} 
            \textbf{a--d} Evolution of the $Q_3$ population as a function of the $Q_2$ frequency ($\omega_{Q_2}$). The x-axis shows the detuning from the theoretical value $\omega_2(m)$, which is given in the main text. From left to right, $m$ increases from 0 to 3, corresponding to fine scans near the first to last bright spots in the first row of Fig.~\ref{fig-figs2}. Simulations in \textbf{a} and \textbf{d} match the experimental configurations in Fig.~2b and Fig.~2c of the main text, showing excellent agreement. 
            \textbf{e--h} Evolution of the $Q_3$ population versus the $Q_1$ frequency ($\omega_{Q_1}$). 
            As $m$ increases from 0 to 3, the spectra in both the upper and lower panels become progressively smoother, indicating increased stability against frequency variations of the qubits. Due to the mirror symmetry in PST, variations of $\omega_{Q_1}$ have the same effect on the evolution as that of $\omega_{Q_3}$. The frequency shifts induced by the couplers are compensated in the simulations.
        }
        \label{fig-figs3}
    \end{figure}
    
    We simulate the solution space of the 3-qubit PST under conditions resembling the experimental environment, including qubit-qubit and qubit-coupler coupling coefficients, pulse parameters, idle frequency assignments and etc. Following the experimental protocol, we scan the parameter space to identify regions that satisfy the PST condition. In the 1$\times$3 qubit chain, we simulate the population of the end qubit $Q_3$ as a function of the frequency detuning $\Delta$ between even and odd qubits and the coupler frequency $\omega_C$, which determines the coupling strengths. The simulated bright spot distributions, shown in Fig.~\ref{fig-figs2}, are in excellent agreement with the experimental results presented in Fig.2a of the main text. 
    
    Near the solutions with $m=0,1,2$ and $3$, we further simulate the evolution spectrum by scanning the frequencies of $Q_1$ (Fig.~\ref{fig-figs3}a--d) and $Q_2$ (Fig.~\ref{fig-figs3}e--h) respectively, with Fig.~\ref{fig-figs3}a and \ref{fig-figs3}d corresponding to the experimental results in Fig. 2b and Fig. 2c of the main text. The simulation results show that as $m$ increases, the evolution spectrum becomes less fragmented and smoother in shape. This indicates that the PST evolution becomes less sensitive to fluctuations in qubit frequencies for larger values of $m$.

    Notably, for a $1\times3$ qubit chain, the time evolution under a Hamiltonian that satisfies the mirror symmetry condition Eq.~\eqref{eq:mirror} can be solved analytically. In the single excitation subspace, the Hamiltonian takes the form
    \begin{align*}
        H_{3} = \begin{pmatrix}
            0 & J & 0 \\
            J & \Delta & J \\
            0 & J & 0
        \end{pmatrix},
    \end{align*}
    where $J$ denotes the coupling strength between $Q_1$--$Q_2$ and $Q_2$--$Q_3$, and $\Delta$ represents the frequency detuning of $Q_2$ relative to $Q_1$ and $Q_3$. The time evolution operator $U_{3}$ can thus be expressed as $U_{3}=e^{-i H_{3} t}=P e^{-i D t} P^{-1}$. Here $P$ and $D$ represent the eigenvectors and eigenvalues of $H_{3}$, respectively. These matrices are explicitly given by 
    \begin{equation*}
        P = \begin{pmatrix}
            1 & -1 & 1 \\
            \frac{\Delta-\sqrt{\Delta^2 + 8J^2}}{2J} & 0 & \frac{\Delta+\sqrt{\Delta^2 + 8J^2}}{2J} \\
            1 & 1 & 1
        \end{pmatrix},
    \end{equation*}
    and 
    \begin{equation*}
        D = \begin{pmatrix}
            \frac{\Delta - \sqrt{\Delta^2 + 8J^2}}{2} &  &  \\
             & 0 &  \\
             &  & \frac{\Delta + \sqrt{\Delta^2 + 8J^2}}{2}
        \end{pmatrix}.
    \end{equation*}
    
    Given an initial state with a single excitation on $Q_1$, the population on each qubit is given by
    \begin{align*}
        P_{Q_1}(\Delta, \Omega, t) &= \frac{1}{4}(\cos\frac{\Omega t}{2} + \cos\frac{\Delta t}{2})^2 + \frac{1}{4}(\frac{\Delta}{\Omega} \sin\frac{\Omega t}{2} + \sin\frac{\Delta t}{2})^2,\\
        P_{Q_2}(J, \Omega, t) &= \frac{4J^2}{\Omega^2} \sin^2\frac{\Omega t}{2},\\
        P_{Q_3}(\Delta, \Omega, t) &= \frac{1}{4}(\cos\frac{\Omega t}{2} - \cos\frac{\Delta t}{2})^2 + \frac{1}{4}(\frac{\Delta}{\Omega} \sin\frac{\Omega t}{2} - \sin\frac{\Delta t}{2})^2.
    \end{align*}
    
    For simplicity, we define $\Omega^2 \equiv \Delta^2 + 8J^2$. When the evolution time is fixed at half-period while scanning $\Delta$ and $J$, the resulting values of $P_{Q_3}$ are consistent with the results shown in Fig.~\ref{fig-figs2} and Fig.~2a in the main text. Similarly, by fixing $J$ and scanning $\Delta$ and $t$, we obtain results that agree with Fig.~\ref{fig-figs3}a--d and Fig.~2b in the main text.

\section{Parameters calibration}
    \begin{figure}
        \centering
        \includegraphics[width=0.9\columnwidth]{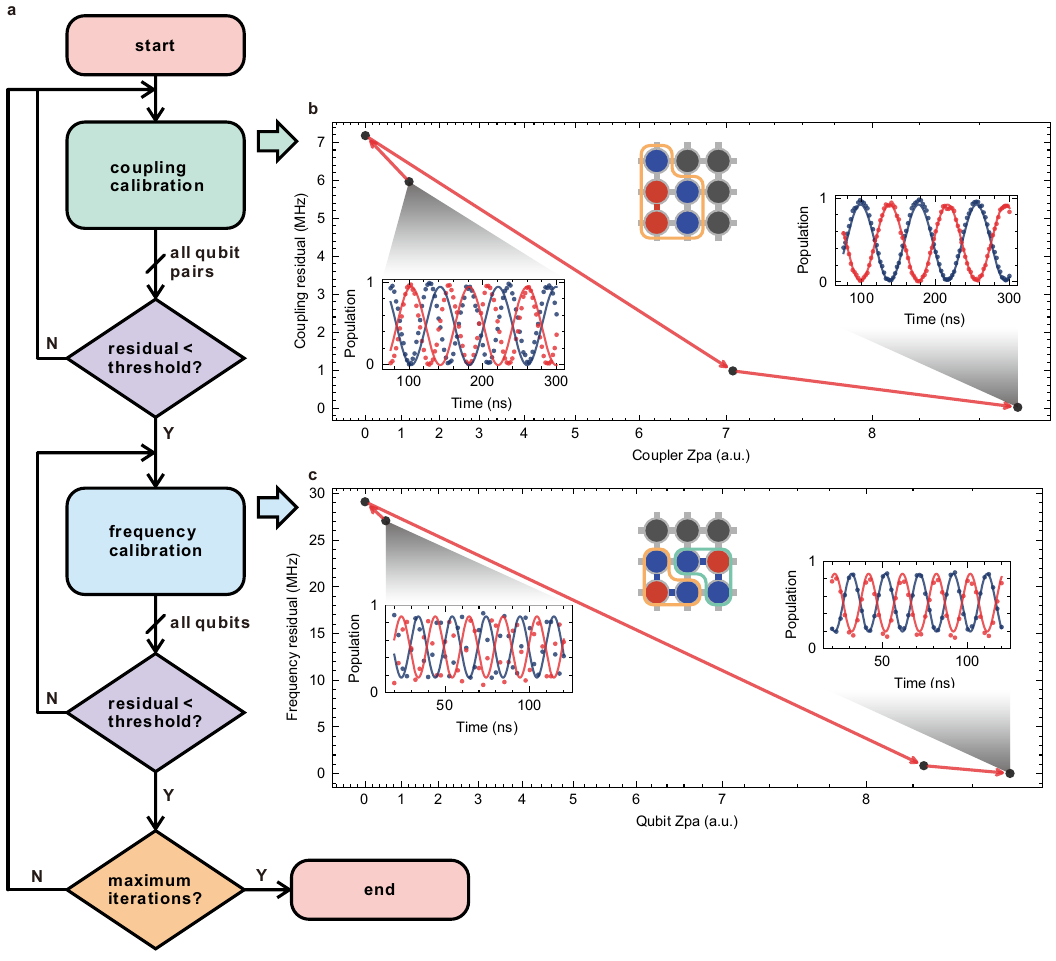}
        \caption{
            \textbf{Parallelizable automated parameter calibration on a $3\times3$ qubit network.} 
            \textbf{a} Flowchart of the parameter calibration procedure. The process begins with the coupling strength calibration. If the residuals between the measured and target coupling strengths for any qubit pair exceeds a predefined threshold (typically 0.1~MHz), recalibration is performed. Once all coupling residuals fall within the threshold, the coupler Zpas are set to the calibrated values, and the process proceeds to the qubit frequency calibration, which also requires all residuals to meet the threshold (typically 0.1~MHz). The above procedures are repeated until a preset maximum number of iterations (typically set to 2) is reached, at which point the entire calibration process is completed.
            \textbf{b} Experimental details of the coupling strength calibration. The calibration unit radius is set to $r=1$, as highlighted by the orange box in the inset. Red qubits and couplers indicate the experimental sites, while blue qubits represent the environmental qubits (see main text for details). The coupling strength is extracted via the swap experiment (shown in the inset). Starting from an initial Zpa value, a custom gradient prediction algorithm iteratively adjusts the Zpa until convergence to the target coupling, as indicated by the trajectory of the black dots and red arrows. Here a.u. is the abbreviation of arbitrary units. 
            \textbf{c} Experimental details of the qubit frequency calibration. With $r=1$ in the $3\times3$ qubit network, two calibration units can be executed in parallel, as illustrated by the orange and green boxes in the inset. The results in the figure correspond to the red qubit within the orange box. Qubit frequency is calibrated through the Ramsey experiment, with the population evolution shown in the subsets. The solid lines represent the target population, and the dots indicate the experimental data.
        }
        \label{fig-figs4}
    \end{figure}
    
    The theoretical model specifies a simple, time-independent Hamiltonian with target qubit frequencies \{$\omega_n$\} and coupling strengths \{$J_{n}$\}. However, it is experimentally challenging to precisely tune these parameters to their target values. This is because \{$\omega_n$\} and \{$J_{n}$\} are tuned by adjusting their Z-pulse amplitudes (Zpas), and the original mapping between these parameters and Zpas is calibrated in an isolated environment, where all other qubits are detuned and all other couplers are turned off. In contrast, our experiments involve multiple qubits and couplers tuned simultaneously to their working points. Such many-body interactions cause the applied Zpas to deviate from the intended frequencies and couplings, resulting in discrepancies in the expected quantum dynamics.

    To address this, we develop a calibration scheme that aims to adjust qubit frequencies and coupling strengths to their targets under conditions as close as possible to the actual experimental environment. To ensure future scalability, the scheme is designed to be both parallelizable and automatable. To enable parallelization, each calibration unit is restricted to a region within radius $r$ around the target qubit or pair, with nearby non-target qubits and couplers defined as environmental sites. For $r=1$, the environmental sites include the nearest-neighbor (NN) qubits of the target qubit and the couplers connecting them. For $r=1.5$, the environmental sites extend to include the outer-layer couplers beyond the NN qubits. For $r=2$, the environmental region additionally includes the next-nearest-neighbor (NNN) qubits. A parallelization threshold $d$ is defined, such that two calibration units are considered parallelizable if the distance between them exceeds $d$. 

    To mitigate the effects of many-body interactions, we employ two distinct frequency configuration schemes for the environmental qubits. The first scheme, as described in Refs.~\cite{2024_param_cali_WangYongYi, 2023_param_cali_LiHao, 2023_param_cali_XiangZhongCheng}, utilizes two sets of mutually symmetric and staggered frequency configurations near the target qubits. Calibration is performed separately for each configuration, and the final result is obtained by averaging the outcomes. However, the asymmetric SQUID design of the qubits limits the available frequency tuning range. As an alternative, we adopt a second scheme in which environmental qubits are set to extreme frequency values -- either the highest or lowest -- far from those of the target qubits. This configuration requires only a single round of calibration, and any accuracy loss can be compensated through subsequent parameter optimization. In practice, the choice between configuration 1 or 2 depends on the specific experimental conditions.

    The flowchart of the automated calibration scheme is illustrated in Fig.~\ref{fig-figs4}a. The calibration experiments are divided into two main categories: frequency calibration for qubits and coupling strength calibration. First, the coupling between two qubits is measured through a swap experiment (Fig.~\ref{fig-figs4}b). The initial Zpa of the coupler is typically set to a value between 1/3 and 1/2 of a period away from its maximum point. A custom gradient prediction algorithm is then used to iteratively determine the Zpa corresponding to the target coupling (usually converging within 5 iterations to an accuracy of 0.1~MHz). Once the Zpas for all couplers are obtained, all couplers are activated. The initial Zpa values for the qubits are then set based on the single-qubit spectroscopy (Fig.~\ref{fig-figs4}c). Qubit frequencies are subsequently measured via the Ramsey experiments. The Zpas corresponding to the target frequencies are then determined using the same gradient prediction algorithm (typically converging within 5 iterations to an accuracy of 0.1~MHz). Finally, by replacing the initial values with those obtained from the calibration process and repeating the above steps once or twice, we obtain the converged calibration values.

    The custom gradient prediction algorithm is explained using the coupling strength calibration as an example. After a swap experiment with an initial Zpa value $Z_0$, the corresponding coupling strength $J_0$ is extracted. Then a small offset is introduced to set the Zpa to $Z_1$, and the new coupling strength $J_1$ is measured. From these two experiments, the gradient is estimated as $k=(Z_1 - Z_0)/(J_1 - J_0)$. This gradient is then used to predict the next Zpa value, $Z_2 = Z_1 + k \cdot (J_{\text{target}} - J_1)$. The data points $(J_1, Z_1)$ and $(J_2, Z_2)$ are subsequently used to update the gradient estimate and predict the next Zpa value. This process is repeated until the coupling strength converges to the target. The Zpa update trajectories for both the coupling strength and qubit frequency are shown in Fig.~\ref{fig-figs4}b and \ref{fig-figs4}c, respectively.

\section{Parameters optimization}
    \begin{figure}
        \centering
        \includegraphics[width=0.9\columnwidth]{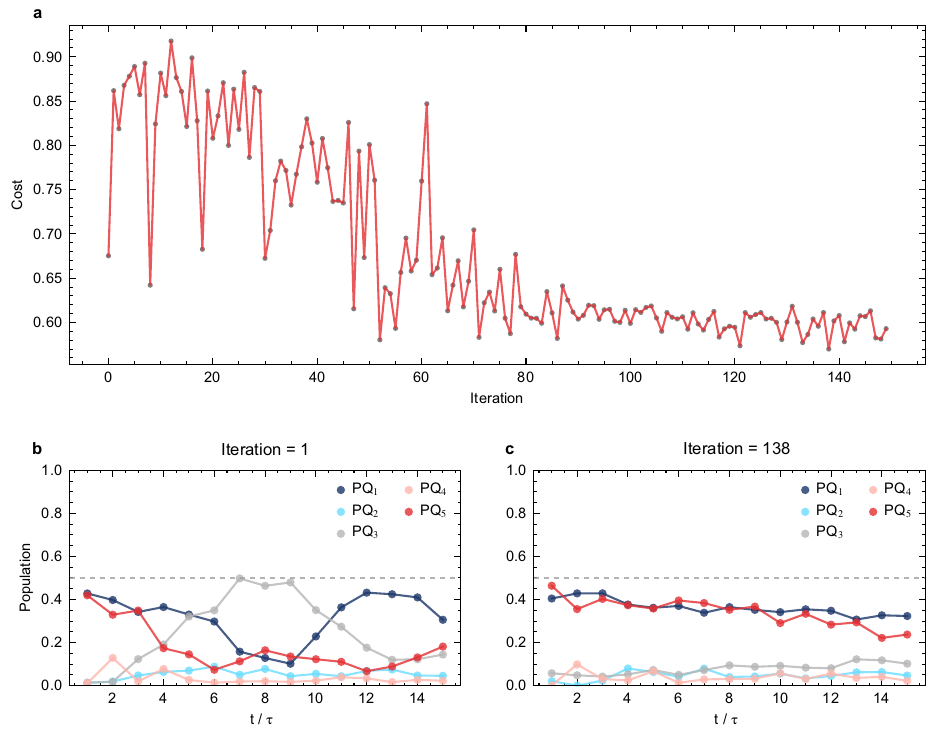}
        \caption{
            \textbf{Nelder-Mead (NM) optimization for the FST experiment on a $1\times5$ qubit chain.}
            \textbf{a} Cost function values as a function of the iteration number. The cost function is defined as $\mathrm{cost} = 1/L \sum_{l=1}^{L} \norm{P_{Q_1}(l) - P_{Q_5}(l)} / \left[P_{Q_1}(l) + P_{Q_5}(l)\right]$, where $P_{Q_n}(l)$ is the population of the $n$-th qubit at the $l$-th sampling point. The number of sampling points $L$ depends on the experimental context. A smaller $L$ is typically used for initial optimization after parameter calibration, while a larger $L$ can be used for further refinement. In this experiment we choose $L=15$. 
            \textbf{b--c} Population distributions at the sampling points before and after optimization, respectively. Ideally, at time $t=n\tau=2\pi n/2J$, the system evolves into a Bell state $(\ket{01} - \ket{10}) / \sqrt{2}$ between $Q_1$ and $Q_5$, leading to the ideal populations $P_{Q_1}^{\text{ideal}}(l) = P_{Q_5}^{\text{ideal}}(l) = 0.5$, as shown by the dashed lines.
        }
        \label{fig-figs5}
    \end{figure}

    \begin{figure}
        \centering
        \includegraphics[width=0.9\columnwidth]{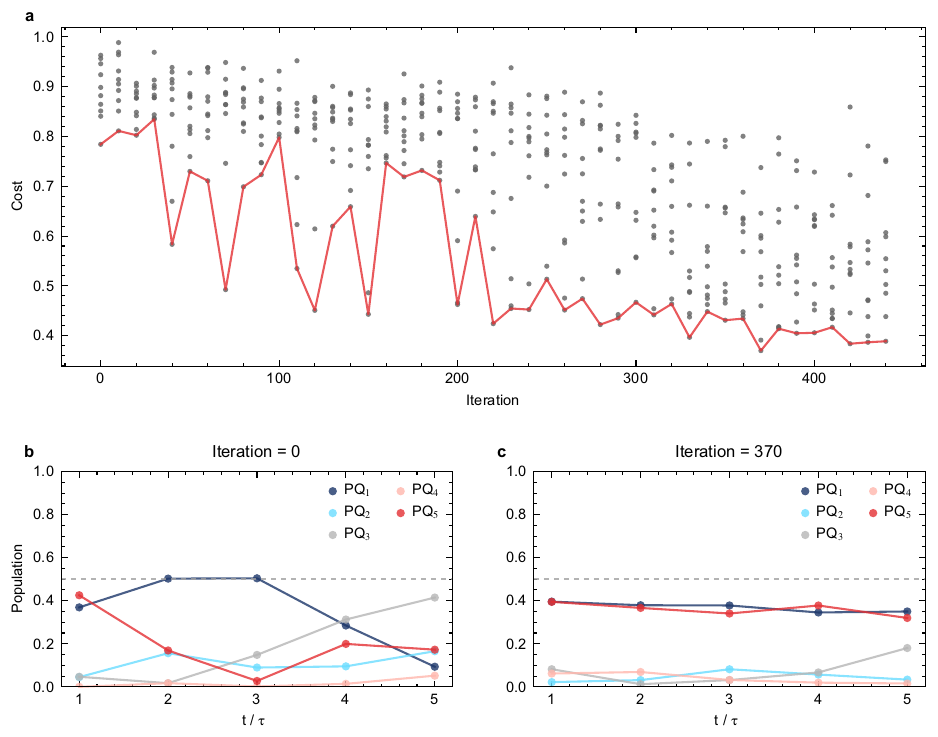}
        \caption{
            \textbf{Differential Evolution (DE) optimization of the FST experiment in a $1\times5$ qubit chain.}
            \textbf{a} Cost function values as a function of the iteration number. The definition of the cost function follows that described in Fig.~\ref{fig-figs5}, with the number of sampling points set to $L=5$ in this experiment. 
            \textbf{b--c} Population distributions before and after the optimization, respectively. The ideal population is indicated by the dashed lines.
        }
        \label{fig-figs6}
    \end{figure}
    
    After estimating the Zpas corresponding to the target frequencies and coupling strengths through parameter calibration, we further optimize the experimental results using a feedback-based optimization algorithm. Depending on the experimental circumstances, we use either a custom Nelder-Mead (NM) algorithm based on the SciPy package, or the Differential Evolution (DE) algorithm available from the GeatPy package. The NM algorithm is a gradient-free simplex method that is robust against gradient fluctuations caused by experimental noise. The DE algorithm, inspired by biological evolution mechanisms, provides stronger global search capabilities and supports larger parameter spaces, albeit at the cost of slower convergence. 

    The ideal population at time $t=n\tau=2\pi n/2J$ ($n=1,2,3,...$) serves as the optimization target, with initial values taken from the parameter calibration procedure. The optimization focuses on the Zpas of all qubits and couplers (with the Zpa of the first qubit fixed since only the frequency difference matters). After several iterations, the algorithm converges to the optimal values. The optimization trajectories of the $1\times5$ qubit FST experiment using the NM and DE algorithms are shown in Fig.~\ref{fig-figs5} and Fig.~\ref{fig-figs6}, respectively. The NM algorithm converges more quickly, with the cost values achieving stable after approximately 100 iterations (Fig.~\ref{fig-figs5}a), while the DE algorithm requires about 250 iterations to converge (Fig.~\ref{fig-figs6}a). Fig.~\ref{fig-figs5}b--c and Fig.~\ref{fig-figs6}b--c compare the population distributions before and after optimization. After parameter optimization, the populations clearly align more closely with the ideal values, indicated by the dashed lines in the figures. The remaining discrepancies from the ideal value can be attributed to decoherence effects or timing misalignment between the experimental sampling points and the theoretical transfer times $n\tau$, which is caused by the rising and falling edges of the pulse.

    It is noteworthy that our optimization schemes are also applicable to PST experiments and those involving 2D cases. We have successfully optimized the parameters for a $3\times3$ qubit network, including the Zpa parameters for 8 qubits and 12 couplers. We believe our optimization schemes has the potential to scale to larger parameter spaces.

\section{Experimental and simulation results of noise robustness}
    \begin{figure}
        \centering
        \includegraphics[width=0.9\columnwidth]{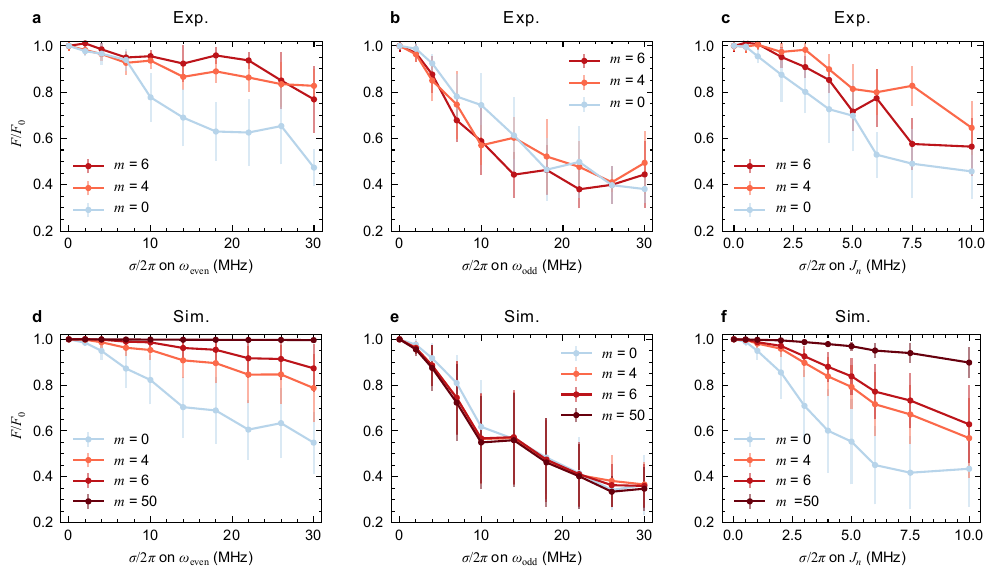}
        \caption{
            \textbf{Robustness of PST fidelity against different types of noise in a $1\times5$ qubit chain}. 
            \textbf{a--c} Experimental results showing the PST fidelity degradation $F/F_0$ under different types of noise: \textbf{a} frequency noise on even-indexed qubits ($\omega_{\text{even}}$), \textbf{b} frequency noise on odd-indexed qubits ($\omega_{\text{odd}}$), \textbf{c} coupling strength noise ($J_n$).
            \textbf{d--f} Corresponding numerical simulation results under the same noise conditions. For each noise intensity, multiple samples are collected (10 in experiments and 50 in simulations). Each data point represents the average fidelity degradation relative to the zero-noise case, with error bars indicating the standard deviation of the fidelities. Fidelity is defined as follows: four distinct initial states $\{ \ket{0}, \ket{1}, (\ket{0}+\ket{1})/\sqrt{2}, (\ket{0}+i\ket{1})/\sqrt{2} \}$ are prepared on the start qubit $Q_1$. After a transfer time $\tau$, the process matrix $\chi_{\text{exp}}$ on the end qubit $Q_5$ is reconstructed through the quantum process tomography. The fidelity is calculated as $\mathrm{Tr}(\chi_{\text{exp}}\chi_{\text{ideal}})$, where the ideal quantum process $\chi_{\text{ideal}}$, from the perspective of $Q_5$, is an identity operation. The Gaussian distributed noise samples are generated using the NumPy random number generator.
        }
        \label{fig-figs7}
    \end{figure}

    \begin{figure}
        \centering
        \includegraphics[width=0.9\columnwidth]{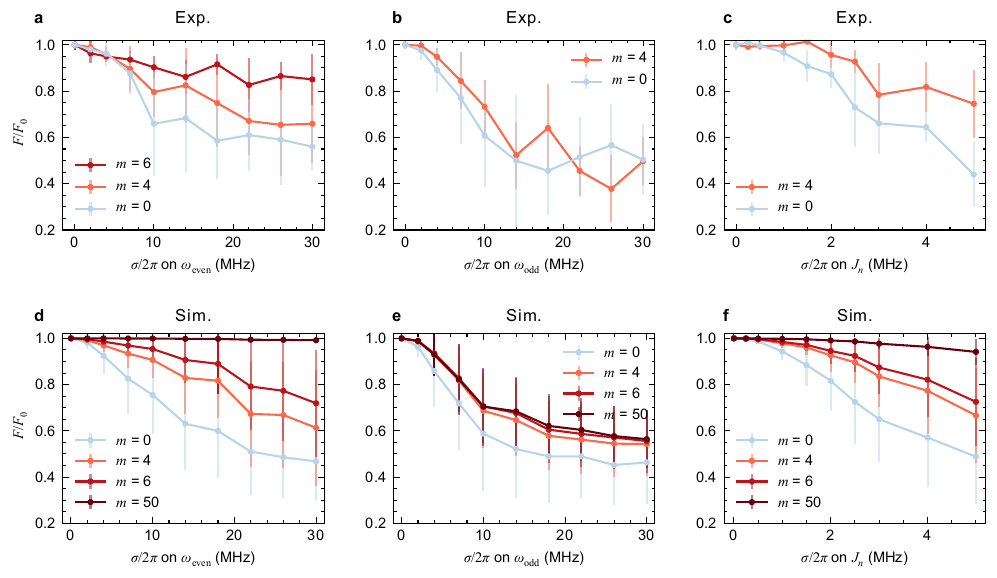}
        \caption{
            \textbf{Robustness of FST fidelity against different types of noise in a $1\times5$ qubit chain}. 
            \textbf{a--c} Experimental results showing the FST fidelity degradation under different types of noise: \textbf{a} frequency noise on even-indexed qubits ($\omega_{\text{even}}$), \textbf{b} frequency noise on odd-indexed qubits ($\omega_{\text{odd}}$), \textbf{c} coupling strength noise ($J_n$). 
            \textbf{d--f} Corresponding numerical simulation results under the same noise conditions. For each noise intensity, multiple samples are collected (10 in experiments and 100 in simulations). Each data point represents the average fidelity degradation relative to the zero-noise case, with error bars indicating the standard deviation of the fidelities. Fidelity is defined as follows: quantum state tomography is performed on $Q_1$ and $Q_5$ to reconstruct their density matrix $\rho_{\text{exp}}$, which is then compared with the ideal density matrix $\rho_{\text{ideal}}$ of the Bell singlet state $(\ket{01} - \ket{10}) / \sqrt{2}$. The fidelity is given by the overlap $\mathrm{Tr}(\rho_{\text{exp}}\rho_{\text{ideal}})$. 
        }
        \label{fig-figs8}
    \end{figure}
    
    In Fig.~\ref{fig-figs7}a--c, we systematically investigate the impact of Gaussian noise on the fidelity of the PST experiments by applying artificial noise to three distinct components: (a) even-indexed qubit frequencies $\omega_{\text{even}}$, (b) odd-qubit frequencies $\omega_{\text{odd}}$ and (c) coupling strengths $J_n$. Fig.~\ref{fig-figs7}d--f show the corresponding numerical simulations under the same conditions, with the additional consideration of a large $m$ ($m=50$) to explore the model's behavior in the extreme regime. Both experimental and simulation results show that the zig-zag configuration with $m>0$ exhibits the highest robustness against the noise on $\omega_{\text{even}}$. Notably, when $m$ is sufficiently large, the fidelity remains nearly unaffected by this type of noise. Furthermore, the model also shows significant robustness to the noise on $J_n$, with the robustness improving as $m$ increases. In contrast, the sensitivity to noise on $\omega_{\text{odd}}$ does not show substantial variations across different $m$ values.

    We also investigate the impact of these three types of noise on the fidelity of FST, as shown in Fig.~\ref{fig-figs8}a--c. The corresponding numerical simulations under the same noise conditions are presented in Fig.~\ref{fig-figs8}d--f, exhibiting a trend similar to the experimental results. The results indicate that the FST fidelity is more resilient to the noise on $\omega_{\text{even}}$ and $J_n$, with fidelity degradation slowing down as $m$ increases. Additionally, under noise on $\omega_{\text{odd}}$, the $m>0$ cases show a slight advantage over the $m=0$ case.

\section{Simulation of the fidelity}
    \begin{figure}
        \centering
        \includegraphics[width=0.9\columnwidth]{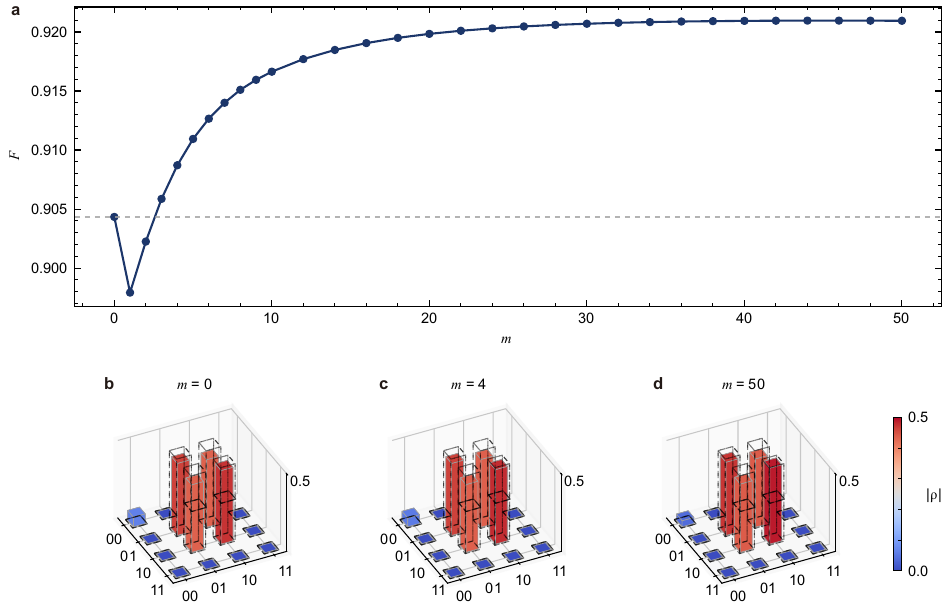}
        \caption{
            \textbf{Simulation of FST fidelity versus $\mathrm{m}$ in a $1\times5$ qubit chain. }
            \textbf{a} Simulated fidelity $F=\mathrm{Tr}(\rho_{\text{sim}}\cdot\rho_{\text{ideal}})$ of remote entanglement generation as a function of $m$, where $\rho_{\text{ideal}}$ represents the density matrix of Bell singlet state $(\ket{01} - \ket{10}) / \sqrt{2}$. The parameter $m$ is scanned from 0 to 50, with $m=0$ corresponding to the line scheme and $m>0$ representing the zig-zag configuration. For each $m$, a master equation simulation is performed, and the fidelity is calculated from the extracted density matrix $\rho_{\text{sim}}$. 
            \textbf{b--d} Simulated density matrices for $m=0$, 4 and 50, respectively. The evolution frequency is set to $J/2\pi = 9$~MHz, consistent with the experimental conditions, yielding an entanglement generation time of $\tau=55.6$~ns.
        }
        \label{fig-figs9}
    \end{figure}

    \begin{figure}
        \centering
        \includegraphics[width=0.9\columnwidth]{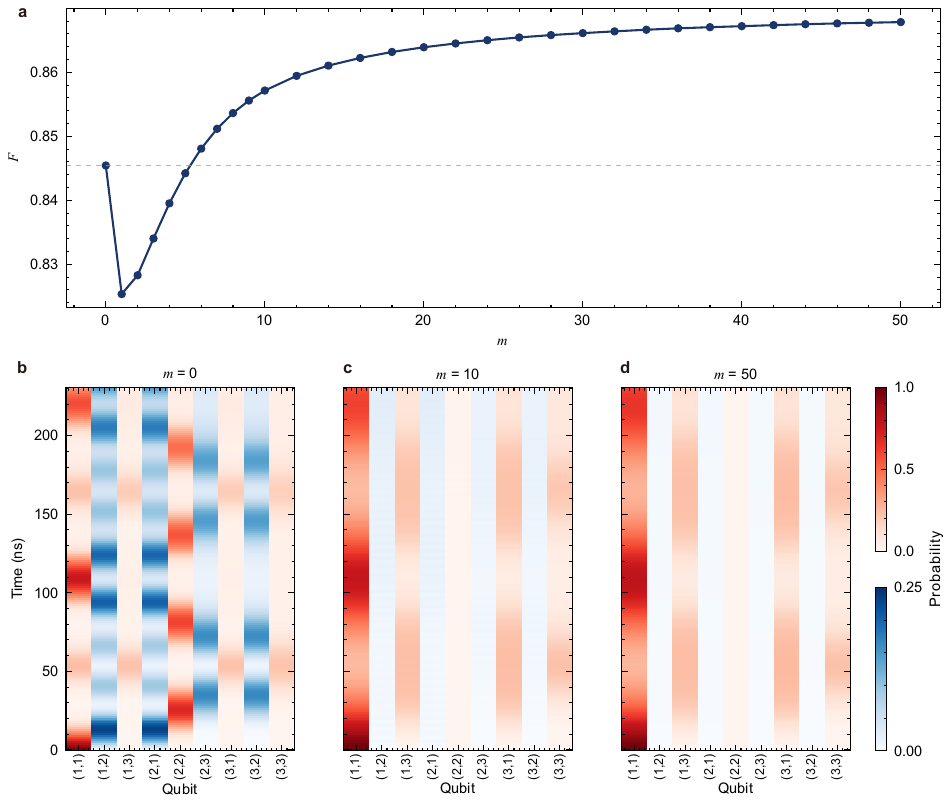}
        \caption{
            \textbf{Simulation of FST fidelity versus $\mathrm{m}$ in a $3\times3$ qubit network. }
            \textbf{a} Simulated fidelity $F=\mathrm{Tr}(\rho_{\text{sim}}\cdot\rho_{\text{ideal}})$ of remote entanglement generation as a function of $m$. $\rho_{\text{ideal}}$ denotes the density matrix of four qubit W state, ${1\over 2}(\ket{1000} + \ket{0100} + \ket{0010} + \ket{0001})$. The parameter $m$ is varied from 0 to 50, with $m=0$ corresponding to the line scheme and $m>0$ representing the zig-zag configuration. For each $m$, a master equation simulation is performed, and the fidelity is calculated from the extracted density matrix $\rho_{\text{sim}}$. 
            \textbf{b--d} Evolution spectra for $m=0$, 10 and 50, respectively. Qubits are labeled by coordinates $(r, c)$, where $r$ and $c$ denote the row and column indices, respectively. The horizontal axis is ordered from left to right according to $n = 3(r-1) + c$, in ascending order. The population of qubits with odd $n$ is shown using a red colormap with a maximum value of 1.0, while that of even $n$ is shown in blue with a maximum value of 0.25. 
        }
        \label{fig-figs10}
    \end{figure}
    
    We analyze the impact of parameter $m$ on fidelity. Numerical simulations are performed using the QuTiP's Lindblad master equation solver, which incorporates both time-dependent Hamiltonian and collapse operators~\cite{qutip_2012_johansson, qutip_2013_johansson}. The simulation employs flattop Gaussian pulses identical to those in the experiments: $\sigma=1.25$~ns, buffer = 7.5~ns for qubits, and $\sigma=2$~ns, buffer = 7.5~ns for couplers. The qubit relaxation times are set to $T_1=16~\mu$s, based on the average value calculated from Table.~\ref{tab:device}. The associated collapse operators are $\sum_n \sqrt{1/T_1} a_n$, where $a_n$ is the annihilation operator of the $n$-th qubit ($n=1,2,3,\dots,N$)~\cite{cqed_blais_2021}. Since the qubit dephasing times $T_2$ at their working points were not measured, we uniformly set $T_2=0.75~\mu$s for 1D qubit chain simulations. The corresponding collapse operators are given by $\sum_n \sqrt{2/T_\phi} a_n^\dagger a_n$, where $a_n^\dagger$ is the creation operator of the $n$-th qubit~\cite{cqed_blais_2021}. Here the $T_\phi$ denotes the pure dephasing time, defined as $1/T_\phi = 1/T_2 - 1/2T_1$.

    The simulation results for remote entanglement generation on a $1\times5$ qubit chain are shown in Fig.~\ref{fig-figs9}. The fidelity is defined as the overlap between the simulated density matrix $\rho_{\text{sim}}$ and the ideal density matrix $\rho_{\text{ideal}}=\kbra{\Psi^-}{\Psi^-}$, where $\ket{\Psi^-} = (\ket{01} - \ket{10}) / \sqrt{2}$ represents the Bell singlet state. Results show that the fidelity first decreases and then increases with increasing $m$, eventually saturating at large $m$ ($m>30$). The saturated fidelity is limited by the overall system decoherence. These results suggest a trade-off point where the zig-zag configuration outperforms the line configuration. Under the parameters in Fig.~\ref{fig-figs9}, this trade-off point occurs at $m=3$. For $m\ge3$, the zig-zag configuration yields higher fidelity than the line configuration. 
    Notably, we find that the position of the trade-off point depends on the pulse configuration. Variations in pulse shapes (e.g., square or flattop Gaussian pulses) and pulse parameters (e.g., $\sigma$ and buffer of the flattop Gaussian pulse) can shift the trad-off point. However, the overall trend of fidelity versus $m$ remains qualitatively consistent. 
    
    Simulations yield the fidelities of 0.910, 0.914 and 0.926 for $m=0, 4$ and 50, respectively. In comparison, the experimental results in Fig.~3 of the main text report fidelities of 0.909 and 0.925 for $m=0$ and $m=4$, respectively. This indicates that the experimental fidelity at $m=4$ nearly matches the saturation fidelity observed in simulation at $m=50$. We attribute this to two main factors. First, the actual pulse may undergo residual distortions during pulse generation and transmission before reaching the qubits, even after pre-distortion calibration, deviating from the ideal pulse used in simulation~\cite{distor_guo_2024}. Second, the working frequencies and coupling strengths (which lead to different levels of frequency shifts on the qubits) differ between $m=0$ and $m=4$. Consequently, in the $m=4$ case, more qubits tend to be closer to their sweet spots, thereby leading to longer $T_2$ times and improved fidelity. 
    
    We further simulate the fidelity of remote entanglement generation in the 2D case as a function of $m$. To reduce computational cost, the Hamiltonian was restricted to the zero- and single- excitation subspace. Specifically, for an $R \times C$ qubit network, the dimension of the simulated Hamiltonian is $(RC+1)\times(RC+1)$, in contrast to the full Hilbert space of dimension $2^N \times 2^N$ used in previous $1 \times N$ chain simulations. In this restricted subspace, the collapse operators for relaxation and dephasing are defined as $\sum_k \sqrt{1/T_1} \ket{0}\bra{k}$ and $\sum_k \sqrt{2/T_\phi} \ket{k}\bra{k}$~\cite{qudit_fischer_2023, decoherence_cai_2006}, respectively, where $k=1,2,\dots, R(r-1) + c,\dots,RC$ indexes the qubits at the $r$-th row and $c$-th column. The relaxation and dephasing times are set to $T_1=16~\mu$s and $T_{\phi}=0.5~\mu$s. The pulse parameters used in these simulations are identical to those in the 1D case. 

    The simulation results in Fig.~\ref{fig-figs10}a demonstrate that the dependence of FST fidelity on $m$ in the 2D case follows a similar trend to that of the 1D qubit chain, with a clear trad-off point. Under the current simulation parameters, this trade-off appears at $m=6$. For $m\ge6$, the zig-zag configuration yield higher fidelity than the line configuration, with the saturation observed beyond $m=50$. Fig.~\ref{fig-figs10}b--d show the evolution spectra of qubit populations for $m=0$, 10 and 50, respectively. As $m$ increases, the population on even-indexed qubits is significantly suppressed, which directly contributes to the improved fidelity. 

\end{document}